\documentclass[sigconf,nonacm]{acmart}

\setcopyright{none}

\usepackage[utf8]{inputenc}
\usepackage{amsmath} % For ampersand in section titles
\usepackage{multirow} % Required for multi-row cells
\usepackage{booktabs}  % For professional-looking table rules 
\usepackage{graphicx}  % For rotating text (\rotatebox)
\usepackage[table]{xcolor} 
\usepackage{tabularx} 
\usepackage{adjustbox} % <-- Added this package
\definecolor{steelblue}{RGB}{70,130,180} % SteelBlue
\definecolor{myorange}{RGB}{255,165,0}   % Orange
\usepackage{graphicx}
\usepackage{booktabs}
\usepackage{tabularx}

\begin{document}
\title[Characterizing User-Reported Risks Across LLM chatbots]{%When the Model Fails: A Large-Scale Empirical Study of User-Reported Risks with Large Language Models
Towards Trustworthy AI: Characterizing User-Reported Risks across LLMs ``In the Wild''}


\settopmatter{authorsperrow=4}

\author{Lingyao Li}
\authornote{Both authors contribute equally to this research, and both are corresponding authors.}
\email{lingyaol@usf.edu}
\affiliation{%
  \institution{University of South Florida}
  \city{Tampa}
  \country{USA}
}

\author{Renkai Ma}
\authornotemark[1]
\email{renkai.ma@uc.edu}
\affiliation{%
  \institution{University of Cincinnati}
  \city{Cincinnati}
  \country{USA}
}

\author{Zhaoqian Xue}
\email{xuezha@pennmedicine.upenn.edu}
\affiliation{%
  \institution{University of Pennsylvania}
  \city{Philadelphia}
  \country{USA}}

\author{Junjie Xiong}
\email{junjiexiong@mst.edu}
\affiliation{%
  \institution{Missouri University of Science and Technology}
  \city{Rolla}
  \country{USA}}



\begin{abstract}
    % While Large Language Models (LLMs) are increasingly integrated into daily life, research on their risks often remains lab-based, disconnected from real-world user experiences. Existing HCI studies often focus on single LLMs like ChatGPT or specific risks like privacy. To gain a holistic understanding of multi-risk across LLM chatbots, we analyze Reddit discussions around seven major LLM chatbots using the NIST AI Risk Management Framework. We find that user-reported risks are unevenly distributed and chatbot-specific. ``Valid and Reliable'' risks are most frequently discussed, but each LLM exhibits a distinct ``risk fingerprint:'' ChatGPT is associated with safety and fairness concerns, Gemini with privacy, and Claude with security and resilience. Less frequent risks, such as explainability and privacy, appear as user trade-offs. Our findings reveal gaps between risks reported by system-centered studies and by users, highlighting the need for user-centered approaches that support users in their daily use of LLM chatbots.

  While Large Language Models (LLMs) are rapidly integrating into daily life, research on their risks often remains lab-based and disconnected from the problems users encounter ``in the wild.'' While recent HCI research has begun to explore these user-facing risks, it typically concentrates on a singular LLM chatbot like ChatGPT or an isolated risk like privacy. To gain a holistic understanding of multi-risk across LLM chatbots, we analyze online discussions on Reddit around seven major LLM chatbots through the U.S. NIST's AI Risk Management Framework. We find that user-reported risks are unevenly distributed and platform-specific. While ``Valid and Reliable'' risk is the most frequently mentioned, each product also exhibits a unique ``risk fingerprint;'' for instance, user discussions associate GPT more with ``Safe'' and ``Fair'' issues, Gemini with ``Privacy,'' and Claude with ``Secure and Resilient'' risks. Furthermore, the nature of these risks differs by their prevalence: less frequent risks like ``Explainability'' and ``Privacy'' manifest as nuanced user trade-offs, more common ones like ``Fairness'' are experienced as direct personal harms. Our findings reveal gaps between risks reported by system-centered studies and by users, highlighting the need for user-centered approaches that support users in their daily use of LLM chatbots.

  %we conduct a large-scale empirical study of user-reported risks, analyzing public discussions on Reddit focusing on seven major LLM chatbots, including GPT, Gemini, Claude, Llama, Qwen, DeepSeek, and Mistral. 
  %Using the NIST AI Risk Management Framework and knowledge graph techniques, we systematically categorize and quantify user-experienced risks. Our findings reveal that user concerns are overwhelmingly dominated by issues of performance, with the ``Valid and Reliable'' category—encompassing hallucinations, inconsistency, and factual errors—constituting the vast reported problems. Beyond this, we find that different LLMs exhibit unique ``risk fingerprints;'' for instance, , suggesting they are less apparent in everyday use. This study contributes a user-centered, empirical characterization of LLM risks, offering actionable insights for designers, developers, and policymakers to prioritize mitigation efforts that align with users' lived experiences.
\end{abstract}








\maketitle

\section{INTRODUCTION}

Large Language Model (LLM), such as OpenAI's ChatGPT, Google's Gemini, and Anthropic's Claude, are rapidly becoming integrated into everyday life~\cite{civitarese2025large}. According to OpenAI COO Brad Lightcap, ChatGPT alone now has seen over 400 million weekly users~\cite{rooney2025openai}. From developers debugging code and writers drafting articles to individuals seeking information and creating content, these tools have moved well beyond research settings~\cite{chen2024large, tian2024debugbench, meyer2023chatgpt}. Their widespread adoption marks a fundamental shift in user experience, where LLM-based conversational agents function as collaborators, assistants, or information brokers~\cite{wu2022ai, jung2024we, walker2024they}. 

While the benefits of LLMs are various~\cite{ziems2024can, li2024scoping, yu2024large}, prior work has also well documented their risks from a system-centered view. These include generating toxic content~\cite{gehman2020realtoxicityprompts}, ``hallucinating'' non-factual information~\cite{ji2023survey}, creating societal biases related to race and gender~\cite{bender2021dangers, gallegos2024bias, kotek2023gender, lee2025investigating}, and security vulnerabilities to adversarial attacks like prompt injection~\cite{shu2025attackeval, xiong2025invisible} and data poisoning~\cite{alber2025medical}. However, much prior work has been conducted in controlled and laboratory-like environments that may fail to capture the unanticipated ways in which risks emerge during the everyday use phase of LLMs. Furthermore, researcher-defined risks of LLMs may not align with what users perceive as risky.

User-facing risks range from user privacy concerns ~\cite{jairoun2024benefit} to biased recommendations in high-stakes domains like healthcare~\cite{templin2025framework}. As LLMs become integral to public use, recent HCI research has begun to investigate key facets of user experience such as model satisfaction~\cite{kim2024understanding}, trust formation~\cite{sun2024trust, jung2024we}, and the trade-offs users face between privacy and utility~\cite{zhang2024s}. However, this line of work has often focused on singular risks like trust and privacy or on a single LLM chatbot, most of the time, ChatGPT. This creates an opportunity to develop a more holistic understanding of the risks in user experience across different LLM chatbots, especially given the increased daily use of multiple chatbots like Gemini, Deepseek, and Claude \cite{ETOnline2025TopAITools, PYMNTS2025DeepSeekAdoption}. To address this gap, we adopt a user-centered approach grounded in large-scale public discourse from Reddit. Specifically, we ask:

\begin{itemize}
    \item \textbf{RQ1}. How can user-reported risks for major LLM chatbots be categorized, and what is the prevalence of each risk type across those chatbots?
    \item \textbf{RQ2}. What are users' lived experiences of navigating these prevalent risks of different LLM chatbots?
\end{itemize}

To answer these questions, we first adopt the U.S. National Institute of Standards and Technology (NIST)’s AI Risk Management Framework (ARF) \cite{ai2023artificial} to bring structure to the diverse spectrum of user-experienced risks. We use this framework as a lens and knowledge graph techniques to analyze large-scale online discussions from Reddit, a well-established HCI method for understanding users' experiences with digital technologies through online data (e.g., \cite{Ma2021, zhang2025dark, chen2025investigating}). 
%finding overview
We find that while performance failures under ``Valid and Reliable'' are the most frequent user-reported risk, each LLM chatbot exhibits a distinct ``risk fingerprint'' with secondary concerns such as safety, privacy, or security (RQ1). We further find that the lived experience of these risks varies, with less frequent risks manifesting as user trade-offs and more common ones experienced as direct, personal harms (RQ2).
%Our analysis reveals that the user-reported risks are overwhelmingly related to model consistency and reliability, such as factual errors and inconsistent performance (RQ1). We then find that different LLM chatbots exhibit unique ``risk fingerprints'' based on their secondary concerns, such as safety, privacy, or security (RQ2). 
%discussion overview
We discuss the disconnect between system-centered and user-centered risk perspectives, how these manifest across different LLM chatbots, and the trade-offs users make. %and the reasons why some technically significant harms remain largely ``invisible'' to users.
Finally, we offer design and policy implications that argue for a truly user-centered model of AI risk management, one grounded in an understanding of users' lived experiences.

Our study makes three main contributions. First, we provide a large-scale, empirical characterization of the user-reported risks across seven major closed- and open-source LLM chatbots, revealing distinct patterns of the risks. Second, we identify three types of trade-offs that users make to navigate these risks in exchange for LLM chatbot utilities, offering a new lens to understand the lived experience of using these chatbots. Finally, we suggest implications for designers and policymakers who advocate for a shift from purely technical risk mitigation to a user-centered approach that supports users in their daily use of LLM chatbots.

%Although the potential of using LLMs to improve human creativity and intellectual productivity is clear, their growing use by the general public, not just AI experts, makes it essential to understand the real-world user experience and the challenges that arise from interactions ``in the wild.''

%their use is accompanied by potential risks. These include the generation of plausible but incorrect outputs (a phenomenon known as ``hallucinations'')~\cite{zhang2025siren}, the amplification of societal biases~\cite{kotek2023gender}, and vulnerabilities to security threats such as prompt injection~\cite{shu2025attackeval, xiong2025invisible} and data poisoning~\cite{alber2025medical}. developing an evidence-based understanding of how users perceive and experience these risks is essential for advancing responsible AI.
%Prior work has examined the technical vulnerabilities of LLMs—highlighting issues such as bias~\cite{gallegos2024bias, lee2025investigating}, adversarial attacks~\cite{shu2025attackeval, xiong2025invisible, qi2024visual}, and alignment failures~\cite{anwar2024foundational}. 

\section{RELATED WORK}
This section reviews prior work on risks of AI and LLMs\footnote{While this paper's focus is on LLM as a prominent class of AI, many of the risks discussed around AI in general are applicable to LLMs, so we use the terms AI and LLM interchangeably.}. We begin by examining technical vulnerabilities as defined through the AI Risk Management Framework, then turn to user-reported risks to motivate our study of multi-risk, cross-chatbot user experiences.

\subsection{Taxonomizing Risks of AI and LLMs: A System-Centered View} 
% Prior research on LLM risks has adopted system-centered methods to investigate technical vulnerabilities, such as harmful content~\cite{gehman2020realtoxicityprompts} and societal biases~\cite{gallegos2024bias}. 

To structure the technical vulnerabilities in AI and LLMs, frameworks like the U.S. National Institute of Standards and Technology’s AI Risk Management Framework (AI RMF) help organizations manage AI risks and promote trustworthy AI development~\cite{ai2023artificial}. It emphasizes managing AI risks across seven dimensions, including \textit{Safe}, \textit{Valid and Reliable}, \textit{Fair}, \textit{Secure and Resilient}, \textit{Accountable and Transparent}, \textit{Explainable and Interpretable}, and \textit{Privacy-Enhanced}.

\textbf{Safe.} The AI RMF defines safety by a system's ability to operate without endangering human life, health, property, or the environment under defined conditions. For example, LLMs pose a safety risk by being able to generate toxic content~\cite{gehman2020realtoxicityprompts}. For instance, prompting LLMs such as ChatGPT with certain personas can systematically elicit toxic or discriminatory responses targeting specific groups, including racial minorities~\cite{deshpande2023toxicity}.

\textbf{Valid \& Reliable.} A valid and reliable AI system can fulfill its intended purpose and perform correctly over time. A core challenge of LLMs is their capacity to generate non-factual content, from unintentional ``hallucinations''~\cite {ji2023survey} to deliberate disinformation~\cite{goldstein2023generative}. Such synthetic content makes it difficult for humans to discern fact from fiction~\cite{clark2021all}, which poses long-term risks to science, education, and social truth~\cite{wachter2024large}. For example, malicious users can intentionally misuse LLMs to automate influence operations, reducing costs while amplifying the scale of propaganda~\cite{bontridder2021role, goldstein2023generative}.

\textbf{Fair---With Harmful Bias Managed.} Fairness involves managing bias to promote equality of AI. Technically, LLMs risk perpetuating unfairness by inheriting biases from their training data, as language models are known to reflect and reproduce societal prejudices related to race and gender~\cite{bender2021dangers}. These biases further manifest as misrepresentation, stereotyping, and derogatory language~\cite{gallegos2024bias}, such as gender biases in job market~\cite{kotek2023gender} and anti-Muslim bias~\cite{abid2021persistent}. %Ultimately, such outputs inequitably represent specific demographics~\cite{albdrani2023investigating}.

\textbf{Secure \& Resilient.} The AI RMF defines a secure and resilient system by its ability to withstand adverse events from unauthorized access. The security of LLMs is a significant technical concern, as these models can be vulnerable to attack and misuse. For example, attackers can prompt LLMs to produce malicious code and reveal security gaps by learning from unresolved vulnerabilities in their training data~\cite{europol2023impact}. Also, perpetrators can misuse AI for crime, such as generating illicit content that is hard for detection~\cite{king2020artificial}.

\textbf{Accountable \& Transparent.} The AI RMF defines transparency by how much information about a system is available to users, which is a prerequisite for accountability. The core problem is the AI model's opacity, where it cannot provide a complete picture of its internal knowledge or reasoning~\cite{bowman2024eight}. This is partly due to the complex transformer architecture, where common interpretation methods can be misleading~\cite{jain2019attention}. 

\textbf{Explainable \& Interpretable.} Explainability and interpretability concerns \textit{how} an AI system works, and interpretability clarifies the \textit{meaning} of its output. Since many machine learning models are intrinsically opaque, these fields aim to illuminate their internal mechanisms or decision processes. While it is well-known that AI is ``black-box''~\cite{doshi2017towards}, a range of methods can (e.g., using input segments to explain AI predictions) help make it more explainable~\cite{belinkov2019analysis}. %Even some research explores using LLMs to generate personalized explanations for other AI systems~\cite{miller2019explanation}.

\textbf{Privacy-Enhanced.} Privacy involves safeguarding human autonomy, identity, and dignity through norms and practices. However, LLMs can memorize and leak sensitive training data guided by specific prompts~\cite{carlini2021extracting}. Consequently, research has focused on technical mitigation like data sanitization and differentially private training~\cite{kandpal2022deduplicating}. New risks also emerge with LLM agents, which could expose sensitive data by clicking on phishing links~\cite{yang2024towards, tang2024prioritizing}.

While this prior work provides a comprehensive understanding of technical vulnerabilities and risk mitigation in AI and LLMs through the system-centered approach, this perspective often overlooks how risks are perceived and experienced by end-users. To address this, we next examine the user-centered perspective on risks that emerge in the later stages of the LLM development cycle, the use phase.

\subsection{Investigating User-Reported LLM Risks}
Prior work across domains has started to identify the risks of using LLM chatbots. For example, users have raised privacy concerns about unauthorized access to their data when interacting with ChatGPT~\cite{alkamli2024understanding}. In education, Harvey et al. classify these risks into two types: technical harms inherent to the AI—such as toxic content, stereotyping, and hallucinations—and broader impacts arising from interactions among students, teachers, and the LLM~\cite{harvey2025don}. Healthcare researchers further highlight that a user's adoption of ChatGPT is contingent on their trust in it~\cite{chen2024perceptions, choudhury2023investigating}. This contingency creates a benefit-risk tension: while LLMs can help patients with personalized education for managing illnesses like diabetes, they also introduce challenges around data security and the need for explainable AI~\cite{jairoun2024benefit}. These risks are especially pronounced in mental health, where clinicians warn of client over-reliance and incorrect treatment recommendations from LLMs~\cite{hipgrave2025balancing}.

Indeed, HCI researchers have been investigating these user-facing risks through several key facets. First, user satisfaction with LLM chatbots like ChatGPT has been a primary focus. While ChatGPT can enhance users’ perception of productivity and accomplishment~\cite{kobiella2024if}, users often report dissatisfaction when it fails to grasp their intent or produces inaccurate responses~\cite{kim2024understanding}. Second, LLMs’ interface design can significantly impact users' trust~\cite{sun2024trust}. However, findings on this are mixed. For example, Walker et al. show that students are cautious and prefer human sources~\cite{walker2024they}, but Yun et al. find that LLM chatbots can elicit higher trust than search engines~\cite{yun2025framing}. Researchers have also explored ways to mediate this trust. For example, Kim et al. find that first-person phrases like ``I'm not sure, but...'' can reduce overreliance on LLM-generated answers~\cite{kim2024m}. Last, HCI researchers have uncovered other risks, such as the trade-offs users face between privacy and utility~\cite{zhang2024s} and the potential for malicious misuse~\cite{li2025closer}. 

% , and users typically prefer other online sources over ChatGPT~\cite{jung2024we}.. revealing that how an LLM expresses uncertainty affects user reliance

However, prior work has largely concentrated on singular risks like trust and privacy or on a single LLM chatbot, oftentimes, ChatGPT. This creates an opportunity to develop a more holistic understanding of the user-reported risks while using different LLM chatbots, especially given the increased daily adoption of chatbots like Gemini, Deepseek, Claude, etc \cite{ETOnline2025TopAITools, PYMNTS2025DeepSeekAdoption}. Such a cross-chatbot and multi-risk investigation is needed to ground policy guidelines, like the AI RMF risk framework, in the real-world user experiences.

\section{METHODS}

\begin{figure*}[htbp]
  \centering
  \includegraphics[width=1\textwidth]{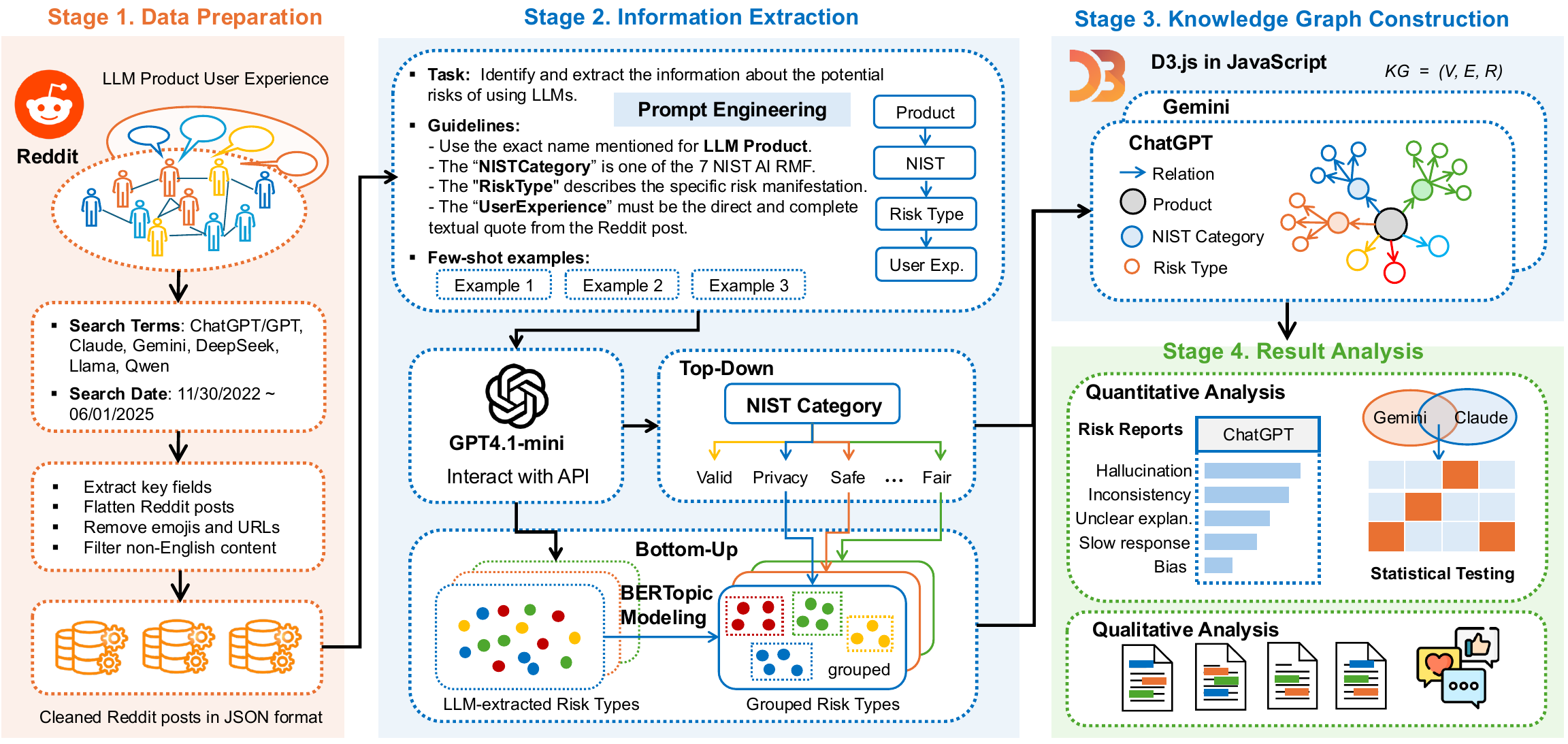}
  % \vspace{-0.35cm}
  \caption{The illustrative framework to implement this study.}
  \label{fig:framework}
\end{figure*}

The study design is illustrated in Figure~\ref{fig:framework}. In Stage 1 (Section~\ref{sec: data}), we gather and filter Reddit data (i.e., both posts and comments) containing terms mentioning major LLM chatbots like ChatGPT and Gemini. In Stage 2 (Sections~\ref{sec: top-down} and~\ref{sec: bottom-up}), we use both top-down and bottom-up methods to extract user-reported risks of using LLM chatbots. In Stage 3 (Section~\ref{sec: kg-build}), we use the extracted information to build a Knowledge Graph visualized with D3.js. This graph lays a foundation for subsequent quantitative and qualitative analyses.

%For the top-down approach, we build a pipeline supported by GPT-4.1-mini; after confirming high inter-rater reliability between human and GPT annotations, this pipeline extracts key dimensions: the specific ``LLM chatbot,'' the high-level ``NIST Risk'' category, a granular ``Risk Type'' that serves an initial code for each user experience and original ``User Experience'' quotations from the data. The bottom-up method then uses BERTopic modeling on the LLM-extracted ``Risk Types'' to group identified risk types. 

\subsection{Data Preparation}
\label{sec: data}

To ground our study in real-world user experiences, we collect a large-scale corpus of public discussions about LLM chatbots from Reddit for two main reasons. First, it is an established practice in prior HCI studies to examine how users discuss their experiences with specific technologies (e.g., \cite{chen2025investigating, Ma2021}). Second, Reddit's forum-style structure supports in-depth, threaded conversations, and its content is often unsolicited and organically generated, capturing more authentic user experiences than those elicited in surveys or interviews. Next, our data collection focuses on seven LLM chatbots, including ChatGPT, Claude, Gemini, Deepseek, Llama, Mistral, and Qwen, selected based on consistent media reports of their high user downloads \cite{ETOnline2025TopAITools, PYMNTS2025DeepSeekAdoption, Bailyn2025MarketShare}. The data collection period spans from November 30, 2022, corresponding with the public release of ChatGPT, the most representative LLM chatbot, through June 1, 2025. To ensure comprehensive coverage of both technical communities and other broader public discussions, we use these seven chatbots as search keywords in a diverse set of 51 subreddits, including 1) seven LLM chatbot-specific communities, 2) 24 AI and LLM-focused, and 3) 20 high-traffic general-interest ones (see Table \ref{tab:datacollection_full}).

\begin{table*}[h!]
\centering
% \small
\caption{Search parameters for the Reddit data collection.}
\label{tab:datacollection_full}
\begin{tabular}{p{3cm} p{14cm}}
\toprule
\textbf{Parameter} & \textbf{Search Conditions} \\
\midrule
\textbf{Search keywords} & ChatGPT, GPT, Claude, Gemini, Deepseek, Llama, Mistral, Qwen \\
\addlinespace
\textbf{Search period} & November 30, 2022 – June 1, 2025 \\
\addlinespace
\textbf{Subreddits} & 
    \textbf{24 LLM-focused:} 
    \newline
    r/AIAssisted, r/AiChatGPT, r/aipromptprogramming, r/Anthropic, r/Chatbots, r/ChatGPT\_Gemini, r/chatgpt\_promptDesign, r/ChatGPT\_Prompts, r/ChatGPTCoding, r/ChatGPTIncreasinglyX, r/ChatGPTPro, r/ChatGPTPromptGenius, r/dalle2, r/GeminiAI, r/GoogleGeminiAI, r/GPTStore, r/LanguageTechnology, r/LLMDevs, r/LocalLLM, r/machinelearningnews, r/OpenAI, r/PromptEngineering, r/SubSimGPT2Interactive, r/SubSimulatorGPT2
    \vspace{0.1em} \newline % Adds a bit of vertical space
    \textbf{20 high-traffic general-interest:} 
    \newline
    r/AskReddit, r/askscience, r/aww, r/books, r/DIY, r/funny, r/gaming, r/Jokes, r/memes, r/movies, r/Music, r/news, r/nottheonion, r/pics, r/science, r/Showerthoughts, r/space, r/todayilearned, r/videos, r/worldnews
    \vspace{0.1em} \newline % Adds a bit of vertical space
    \textbf{7 LLM chatbot-specific:}
    \newline
    r/Bard, r/ChatGPT, r/ClaudeAI, r/DeepSeek, r/LocalLLaMA, r/MistralAI, r/Qwen\_AI \\
\bottomrule
\end{tabular}
\end{table*} 

Next, we collect Reddit data using a custom script built with the Python Reddit API Wrapper (PRAW)~\cite{praw} to scrape top-ranking posts and their comments from selected subreddits with minimal metadata fetched for anonymity. For each post, we fetch the title, body (i.e., post content), creation time, and number of comments. For each comment, we retrieve the parent ID, link ID, body (i.e., comment), %author, score, 
creation time, and associated replies. This initial collection yields a raw dataset of 28,058 posts and 348,231 comments. The raw corpus then undergoes a multi-stage filtering process to ensure data quality and relevance. First, to maintain relevance, we retain only posts with titles or bodies containing at least one of the listed LLM chatbots. Next, we remove duplicate posts and comments based on the content and post time. Last, to enhance data quality, we keep only comments containing more than 10 words, which is important for filtering out low-effort or non-substantive replies (e.g., ``lol,'' ``agree,'' ``thanks'') and improving the signal-to-noise ratio. Finally, we combine the title and filtered body of each post into a single JSONL file to preserve complete context. This process yields a final dataset of 4,438 posts and 48,797 comments for analysis.

\subsection{Top-down Method for Risk Extraction}
\label{sec: top-down}

We use the U.S. NIST AI RMF’s seven characteristics of trustworthy AI~\cite{ai2023artificial} as a top-down analytical lens to interpret user-reported risks, enabling direct mapping onto an established framework. Building on this, we develop an LLM-driven pipeline to automatically extract four analytical dimensions. The pipeline is guided by a prompt consisting of: (1) a system role framing the model as an ``expert in information extraction,'' and (2) a task prompt leveraging in-context learning with few-shot examples that illustrate how to extract structured JSON outputs across diverse scenarios from Reddit posts. To ensure consistency, the prompt encourages implicit chain-of-thought reasoning and provides explicit constraints, such as requiring outputs to use only the seven NIST-defined categories. During execution, Reddit data and the structured prompt are passed to the pipeline, which generates JSON-formatted analyses. A validation function then parses these outputs to verify adherence to the four analytical dimensions (see examples in Appendix \ref{sec:annotation_example}).

%This top level of our framework defines the primary risk categories. These include, (1) Valid and Reliable, (2) Safe, (3) Secure and Resilient, (4) Accountable and Transparent, (5) Explainable and Interpretable, (6) Privacy-Enhanced, and (7) Fair (see specific explanation of these categories in the appendix).

\begin{itemize}
    \item LLM chatbot: This identifies the specific chatbot mentioned in our data, such as \textit{GPT}, \textit{Claude}, or \textit{Llama}.
    \item NIST Category: This classifies the observation into one of the seven risk classifications defined by the AI RMF, which provides a high-level categorization for user-reported risks.
    \item Risk Type: This provides a granular, lower-level risk manifestation, such as ``Data transfer to foreign servers leading to potential privacy concerns,'' or ``Data exposure and loss of data ownership.'' This allows for the identification of specific issues (i.e., initial codes) within each NIST category.
    \item User Experience: This captures the direct and complete textual quote from our data that serves as evidence for RiskType and NISTCategory.
\end{itemize}

 %Specifically, the model is asked to (1) identify a potential risk, (2) classify it using one of the official NIST risk categories, (3) define a more granular RiskType, and (4) extract a direct quote from the Reddit post to serve as evidence. This structured workflow encourages a step-by-step analytical process. 

To validate the pipeline’s output, we construct a ground-truth dataset by randomly sampling 200 entries from the collected Reddit data. The authors are divided into two teams with each assigned 100 samples. Within each team, authors independently extract and inductively code the four analytical dimensions. We then measure inter-coder reliability using Krippendorff’s alpha, obtaining a strong average agreement of $\alpha = 0.82$. After resolving disagreements, we produce a consensus coding for all 200 samples, which yields a reliable ground-truth dataset.

To apply the four analytical dimensions at scale, we first select GPT-4.1-mini as the backbone of the pipeline, given its strong balance of performance, cost-efficiency, and ability to generate reliable structured outputs. Although more advanced models such as GPT-4.1 or o3 may offer slightly higher accuracy, their substantially greater computational cost makes them impractical for large-scale annotation on our dataset. We evaluate the pipeline against the 200 consensus-based ground-truth labels. For the NISTCategory classification, the pipeline achieves an accuracy of 0.87 and a Krippendorff's alpha of $\alpha = 0.76$. For the more descriptive Risk Type dimension, we assess the textual similarity using Recall-Oriented Understudy for Gisting Evaluation, achieving an F1-score of 0.83, indicating strong agreement with human expert judgment. With this validation, we then deploy the pipeline on the full dataset.

\subsection{Bottom-up Method with Topic Modeling for Risk Type's Thematic Grouping}
\label{sec: bottom-up}

To complement the top-down method, we also employ a bottom-up topic clustering approach to identify emergent sub-risk themes directly from the granular ``Risk Type'' descriptions that are identified by our LLM pipeline. Using an embedding-based method, we group semantically similar risk descriptions, which provides a more nuanced understanding than simple keyword matching. Our process utilizes the BERTopic framework~\cite{grootendorst2022bertopic}. Each ``Risk Type'' is first converted into a numerical vector using the \textit{all-mpnet-base-v2} SentenceTransformer model~\cite{all-mpnet-base-v2}. To reduce the complexity of these high-dimensional vectors and improve the effectiveness of clustering, we then employ a dimensionality reduction technique, Uniform Manifold Approximation and Projection (UMAP)~\cite{mcinnes2018umap}, to condense the vectors into a lower-dimensional space while preserving their important semantic features.

After dimensionality reduction, we cluster the vectors to group similar risk types. We apply the elbow method to determine the optimal number of clusters, which finds a balance between topic granularity and interpretability. We then use the K-Means clustering algorithm~\cite{likas2003global} to partition the data into the predefined number of clusters. This method assigns each risk description to the nearest cluster center, grouping related concepts together. To interpret the content of these clusters, we process the text with a CountVectorizer to calculate word frequencies while removing common English stop words. We then apply the class-based Term Frequency-Inverse Document Frequency (c-TF-IDF) transformer~\cite{grootendorst2022bertopic} to identify keywords that are most representative of each topic. 

In the final step, two authors independently review the grouped topics and assign thematic labels based on the representative words for each topic. This thematic scheme is then discussed with all co-authors to reach consensus, ensuring that the themes are coherent and accurately reflect the underlying Risk Type and User Experience quotations. The details of this annotation results are presented in Appendix~\ref{app:topic}.

\subsection{Knowledge Graph Development}
\label{sec: kg-build}
We construct an interactive knowledge graph (KG) to systematically represent multi-level data relationships, empowering HCI researchers and users to explore risks from high-level patterns to individual experiences and facilitating a mixed-methods inquiry. The KG is formally defined as a graph $G = (V, E)$, where $V$ is a set of vertices (nodes) and $E$ is a set of edges. The vertices are partitioned into three distinct types: $V = L \cup C \cup R$ where $L$ represents the set of ``LLM chatbots,'' $C$ the ``NIST Categories,'' and $R$ the granular ``Risk Type'' identified from our bottom-up analysis. The relationships are represented as directed edges to reflect the hierarchical structure (from specific risks to categories to chatbots), where edges form two key relational sets: (1) from Risk Types to NIST Categories, and (2) from NIST Categories to LLM chatbots. Each edge $e$ is weighted by co-occurrence frequency and defined as $e = (source, target, weight)$. Here, $source$ and $target$ are nodes from adjacent layers (e.g., $source \in R$, $target \in C$ or $source \in C$, $target \in L$), and $weight$ is the number of mentions in the dataset. 

% Unlike some KGs that include additional properties (e.g., severity), our focus is on frequency to prioritize prevalence analysis, enhancing reproducibility for HCI explorations.

For the implementation, we utilize the D3.js library to build the KG as an interactive, web-based visualization~\cite{bostock2011d3}. The visualization employs a force-directed layout algorithm to optimize node arrangement and minimize edge crossings, resulting in an intuitive structure. We assign distinct visual encodings to different node types. To ensure clarity across a wide distribution of frequencies, the radius of each node $n$ and the thickness of each edge $e$ are rendered using log-normalized scaling. Specifically, $\text{radius}(n) = \text{base\_size} + \log(\text{frequency}(n))$ and $\text{thickness}(e) = \text{base\_thickness} \cdot \log(\text{weight}(e))$. Here, $\log$ denotes the natural logarithm, with example defaults like $\text{base\_size} = 5$ and $\text{base\_thickness} = 1$ (adjustable in code). This scaling ensures that less frequent nodes remain visible while prominent ones are emphasized.
% supporting perceptually effective visual ranges in HCI contexts.

As an analytical tool, the visualization allows users to filter the graph to isolate areas of interest, and hovering over an edge reveals a data-rich tooltip with direct ``User Experience'' quotes from the source Reddit posts. This mechanism pulls evidence directly from the extracted User Experience dimension, bridging the quantitative overview with rich qualitative data. By enabling deeper, evidence-based understanding of how users articulate and experience specific LLM-related risks, the KG supports our RQ2 on users' lived experiences with these risks.

\subsection{Ethics Statement}
Our study, which utilizes publicly available Reddit discussions, is designed with careful consideration for ethical research practices. Our approach aligns with the common interpretation among Institutional Review Boards (IRBs) that research using publicly available and pseudo-anonymous data is exempt from human subjects review, particularly when it does not involve sensitive or private information \cite{proferes2021studying}. Besides, we are cognizant of the ongoing ethical discourse within the HCI community regarding the use of public online data, with a focus on privacy, anonymity, and the potential risks of identifiability \cite{fiesler2019ethical}. Therefore, to proactively address these concerns and protect the individuals whose discussions inform our work, we implement three protective measures in our study. First, we anonymize the dataset by removing all identifying information before data analysis. Second, all quotations presented in this paper have been paraphrased to reduce their searchability and prevent them from being traced back to the original Reddit posts. Finally, the entire dataset is stored securely on a password-protected device, with access restricted solely to the research team.

\section{RQ1 RESULTS: Prevalence of User-Reported Risks across LLM Chatbots}
\subsection{Knowledge Graph of User-Reported Risks}
\label{kg_finding}

\begin{figure*}[htbp]
  \centering
  \includegraphics[width=0.8\textwidth]{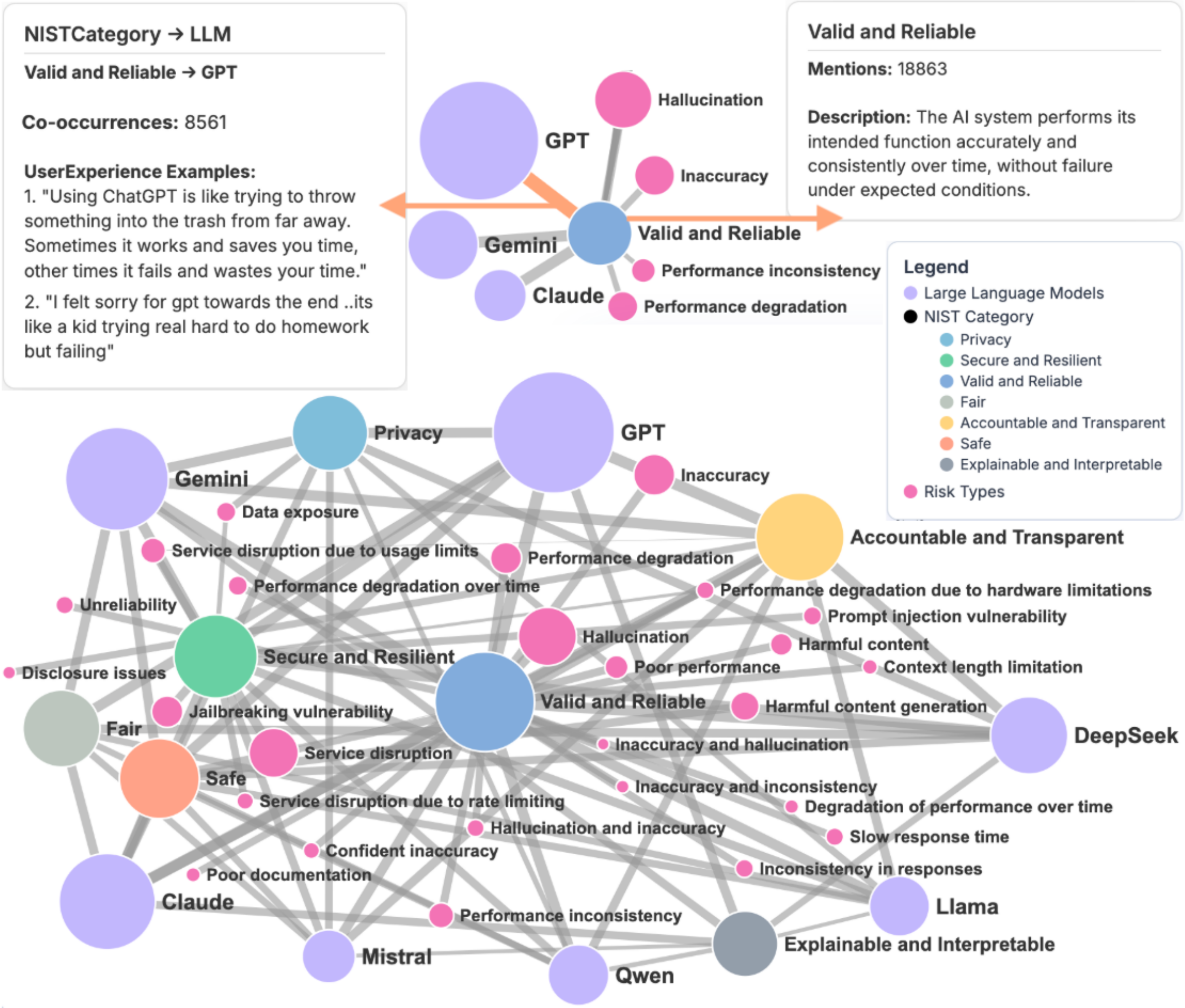}
  % \vspace{-0.35cm}
  \caption{Interactive KG visualization of user-reported LLM risks extracted from Reddit data. (a) The hierarchical relationships between LLM chatbots, NIST's AI RMF categories, and granular risk types, with node sizes proportional to the frequency of mentions. (b) Focused view of the KG for ``Valid and Reliable,'' showing detailed statistics on frequency distributions. Different colors represent different node types. The interactive KG is publicly accessible at \url{https://shorturl.at/DDPi5}.}
  \label{fig:kg}
\end{figure*}

The constructed KG visualizes the interaction network between LLM chatbots and their associated risks reported by Reddit users (Figure~\ref{fig:kg}). The graph represents 32,302 relations connecting 7 LLM chatbot entities with 7 NIST category entities and 25,938 unique granular risk type entities. The graph reveals that ``Valid and Reliable'' is the most frequently reported NIST category, which is visually represented by its significantly larger node size than other categories. The connections between chatbots like GPT and prevalent risk types such as ``hallucinations'' appear substantially thicker than connections to less frequently reported risk types, reflecting the weight of these relationships in discussions.

The interactive capabilities of our implementation (Figure~\ref{fig:kg}B) enable a more detailed examination of specific elements. For instance, selecting ``Valid and Reliable'' automatically generates a focused KG displaying the connections between all LLM chatbot entities and the ``Valid and Reliable'' node. When hovering over a NIST category node such as ``Valid and Reliable'' (Figure~\ref{fig:kg}(b), a tooltip displays comprehensive statistics, including total mentions (18,863) with distributions across LLM chatbots and various other descriptors. Edge tooltips provide detailed insights into specific LLM-NIST category relationships (Figure~\ref{fig:kg}B). The ``Valid and Reliable'' relationship of ChatGPT displays co-occurrences (8,561) and representative user experiences. For instance, the tooltip shows two user examples with varying experiences: one user described, ``Using GPT is like trying to throw something into the trash from far away. Sometimes it works and saves you time, other times it fails and wastes your time,'' highlighting performance inconsistency, while another noted, ``I felt sorry for gpt towards the end..its like a kid trying real hard to do homework but failing.'' 

% Our interactive KG is publicly accessible at \url{https://shorturl.at/DDPi5}.

\subsection{Statistical Analysis of User-Reported Risks}
\label{stats_finding}

\begin{figure*}[htbp]
  \centering
  \includegraphics[width=1\textwidth]{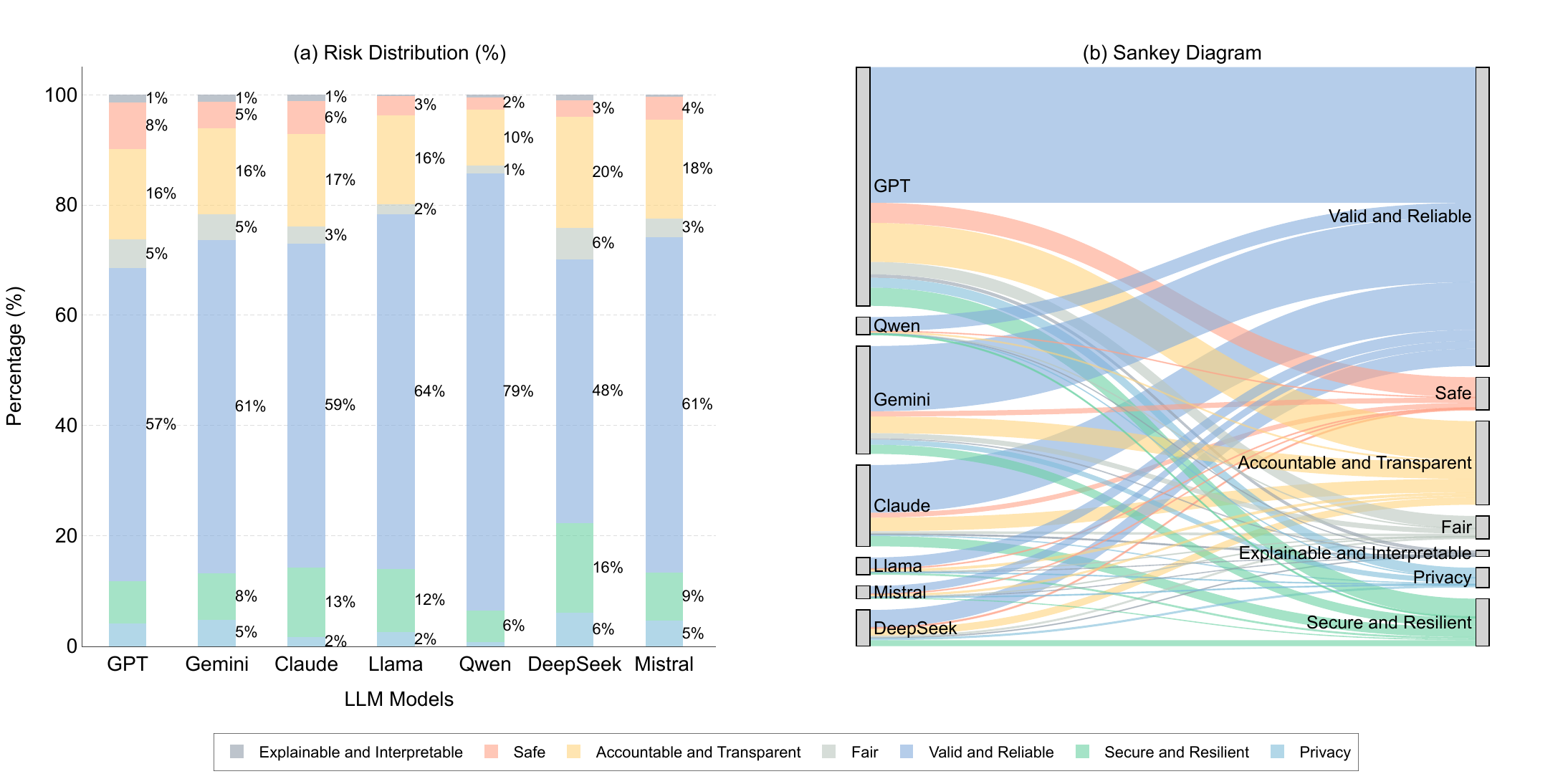}
  \caption{The distribution of user-reported LLM risks. (a) A stacked bar chart showing the percentage breakdown of reported risks across the seven NIST categories for each LLM chatbot. (b) A Sankey diagram illustrating the volume of risk reports flowing from each LLM chatbot (left) to the corresponding NIST categories (right), where the width of the flow is proportional to the number of mentions.}
  \label{fig:distribution}
\end{figure*}

Building on the KG's mapping of risk co-occurrences, we quantify the prevalence of these risks across LLM chatbots. Figure~\ref{fig:distribution}(a) presents the percentage distribution of user-reported risks, categorized according to the NIST taxonomy. This reveals that ``Valid and Reliable'' constitutes the vast majority of user experience (an average of 61.3\%), followed by ``Accountable and Transparent'' (an average of 16.1\%) and ``Secure and Resilient'' (an average of 10.1\%). In contrast, governance-oriented domains such as ``Fair,'' ``Privacy,'' and ``Explainable and Interpretable'' receive significantly less attention. This distribution suggests that users are most likely to report concerns that are immediately apparent in daily use of LLM chatbots, such as factual errors or service disruptions, whereas negative experiences in transparency or fairness may be less obvious or require specific contexts to emerge.

The Sankey diagram in Figure~\ref{fig:distribution}(b) further illustrates these relationships, mapping the flow of reported risks from each LLM chatbot to the NIST categories. The width of each flow is proportional to the volume of reports, visually confirming that ``Valid and Reliable'' is the primary area for user concerns across all models. However, the diagram also highlights distinctive patterns. For example, a substantial secondary flow emanates from GPT to the ``Safe'' category, a pattern less pronounced in other models. Similarly, DeepSeek shows a uniquely strong connection to ``Accountable and Transparent,'' while Claude directs a comparatively large volume of its reported risks to ``Secure and Resilient.''

\begin{figure*}[htbp]
  \centering
  \includegraphics[width=1\textwidth]{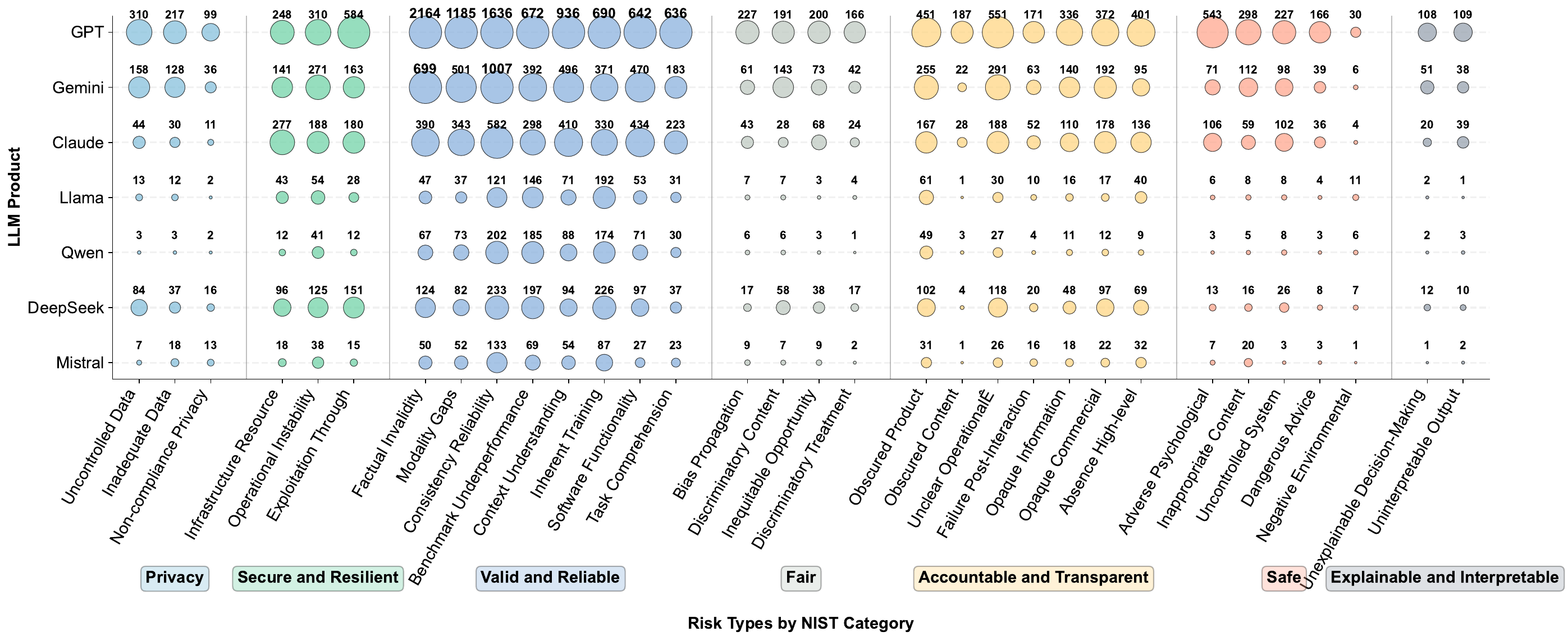}
  \caption{A categorical analysis of granular user-reported risks. The dot plot visualizes the frequency of specific risk sub-categories for each LLM. The size of each dot corresponds to the number of reports for that sub-category.}
  \label{fig:category}
\end{figure*}

To move beyond this aggregate view and understand model-specific issues, Figure~\ref{fig:category} disaggregates these patterns, revealing that each LLM exhibits a unique risk fingerprint. While ``Valid and Reliable'' issues (e.g.. ``'Consistency and Reliability,'' ``Factual Invalidity'') are the most frequently reported concerns for all LLM chatbots, their secondary risk profiles differ. Among the closed-source models, GPT's profile is skewed towards ``Safe'' concerns; users report issues related to ``Adverse Psychological and Behavioral Influence'' and ``Inappropriate Content Generation'' more frequently. Gemini's reported risks also show a distinct focus on ``Privacy,'' where concerns over ``Uncontrolled Data Access and Exposure'' are more prominent. In contrast, Claude presents a highly specialized risk profile: users report a high degree of issues under the ``Secure and Resilient'' category, particularly related to ``Operational Instability and Disruption,'' while showing comparatively few concerns about ``Fairness'' and ``Privacy.''

Among the open-source models, DeepSeek has seen significantly more risk reports. Compared to other models, its user-reported issues are more evenly distributed across multiple categories, with notable discussions in ``Accountable and Transparent,'' ``Secure and Resilient,'' and ``Privacy,'' in addition to common reliability concerns. In contrast, discussions surrounding other open-source models like Llama, Qwen, and Mistral are more heavily dominated by fundamental ``Valid and Reliable'' concerns. Both Llama's and Mistral's profiles are heavily weighted toward issues of ``Factual Invalidity'' and ``Consistency Reliability.'' This pattern is most extreme with Qwen, whose risk profile is the most concentrated on ``Valid and Reliable'' issues of all models, with an overwhelming user focus on its reliability and under-performance but comparatively few complaints across other categories.

Next, we employ a Chi-Square test of independence on the contingency table of \{LLM\}$\times$\{NIST category\}. After confirming that the risk distributions differ significantly across models, we conduct a post-hoc analysis using adjusted standardized residuals to identify which specific categories contribute to this effect. These residuals approximate a normal distribution ($N(0,1)$), where a large magnitude ($|r| \geq 2$, corresponding to a two-sided $\alpha \approx 0.05$) indicates that a risk is reported significantly more (positive residual) or less (negative residual) frequently than expected by chance. As shown in Table~\ref{tab:chi_matrix}, we identify these deviations as ``significantly more'' or ``significantly less.'' It is important to note that these results should be interpreted as directional diagnostics rather than causal statements.

The statistical analysis in Table~\ref{tab:chi_matrix} corroborates the visual patterns, revealing uneven and product-specific user concerns. Among the closed-source LLM chatbots, GPT exhibits significantly more user-facing issues in the ``Safe,'' ``Fair,'' and ``Explainable and Interpretable'' categories, while having significantly fewer complaints regarding technical robustness (``Secure and Resilient'' and ``Valid and Reliable''). Claude presents a near-opposite profile, with more pronounced reports in ``Secure and Resilient'' that is coupled with significantly less in the ``Fair'' and ``Privacy'' domains. Gemini demonstrates a mixed profile, overrepresented in reports concerning ``Privacy'' and ``Valid and Reliable'' but underrepresented in ``Safe'' and ``Secure and Resilient'' issues.

The open-source LLM chatbots exhibit variations. Llama shows a clear focus on technical performance, with significantly more reports than expected in ``Secure and Resilient'' and ``Valid and Reliable'' but consistent deficits across user-facing domains. This specialization is even more extreme in Qwen, which displays a massive positive deviation for ``Valid and Reliable'' ($r = 9.12$) along with significant negative deviations in nearly all other categories. In contrast, DeepSeek shows the most versatile risk profile, with significantly more risk reports spanning ``Accountable and Transparent,'' ``Fair,'' ``Privacy,'' and ``Security and Resilient.'' These are complemented by significantly fewer reports than expected in the ``Safe'' and ``Valid and Reliable'' categories. Finally, Mistral, despite a lower report volume, shows significantly fewer user reports in ``Explainable and Interpretable'' and ``Safe.''

\begin{table*}[t]
\centering
\small
\caption{Post-hoc analysis of Chi-Square test with Pearson residuals for each LLM chatbot and AI RMF risk category pair.}
\label{tab:chi_matrix}

\newcommand{\sigmore}[1]{\cellcolor{steelblue!25}\textbf{#1}}
\newcommand{\sigless}[1]{\cellcolor{myorange!20}\textbf{#1}}
\newcolumntype{Y}{>{\centering\arraybackslash}X}

\begin{tabularx}{\textwidth}{l YYYYYYY}
\toprule
& \multicolumn{7}{c}{\textbf{NIST's AI RMF Risk Category}} \\
\cmidrule(lr){2-8}
\textbf{LLM} & Privacy & Sec. \& Res. & Val. \& Rel. & Fair & Acc. \& Trans. & Safe & Expl. \& Int. \\
\midrule
\multicolumn{8}{l}{\textit{Closed-Source Models}} \\
ChatGPT & \sigmore{1.93} & \sigless{-6.81} & \sigless{-2.51} & \sigmore{3.95} & 0.14 & \sigmore{9.74} & \sigmore{2.31} \\
Gemini & \sigmore{3.71} & \sigless{-2.24} & \sigmore{2.28} & 0.64 & -1.64 & \sigless{-5.22} & 0.56 \\
Claude & \sigless{-8.00} & \sigmore{7.78} & 0.28 & \sigless{-4.52} & 0.72 & -1.13 & -0.53 \\
\midrule
\multicolumn{8}{l}{\textit{Open-Source Models}} \\
Llama & \sigless{-2.29} & \sigmore{2.42} & \sigmore{2.53} & \sigless{-4.01} & -0.19 & \sigless{-3.89} & \sigless{-2.84} \\
Qwen & \sigless{-5.36} & \sigless{-3.84} & \sigmore{9.12} & \sigless{-4.88} & \sigless{-5.07} & \sigless{-5.52} & \sigless{-2.38} \\
DeepSeek & \sigmore{5.27} & \sigmore{11.06} & \sigless{-6.60} & \sigmore{2.66} & \sigmore{4.43} & \sigless{-6.26} & -1.15 \\
Mistral & 1.19 & -0.52 & 0.90 & -1.61 & 1.12 & \sigless{-2.50} & \sigless{-2.22} \\
\bottomrule
\end{tabularx}

% --- Table Note ---
\begin{flushleft}
\vspace{1ex}
\footnotesize
Note: \textcolor{steelblue}{Blue} cells indicate significantly more reports than expected ($|r|>1.96$), while \textcolor{myorange}{Orange} cells indicate significantly fewer. Header abbreviations: Sec. \& Res. (Secure \& Resilient), Val. \& Rel. (Valid \& Reliable), Acc. \& Trans. (Accountable \& Transparent), Expl. \& Int. (Explainable \& Interpretable).
\end{flushleft}
% --------------------

\end{table*}

\section{RQ2 RESULTS: The Lived Experience of Navigating LLM Risks}
This section details the risks of users' lived experiences, organized from the least to the most frequently. In the less frequent ones, we uncover trade-offs, such as users balancing the need for ``Explainability'' against usability or confronting manipulative designs in ``Privacy'' that force them to sacrifice core functionality. More prevalent risks are direct personal and social harms, including emotional harm from stereotyping in Fairness and adverse psychological effects in Safety. These escalate to the most frequently reported risks, including a lack of ``Accountability \& Transparency'' and ``Validity \& Reliability''. By purposively sampling these narratives, we highlight users' lived experiences behind these risks. Full details of all identified themes are available in Appendix \ref{app:topic}.

% many underrepresented in existing technical literature (e.g., \cite{gehman2020realtoxicityprompts, goldstein2023generative}). 

% Explainability \& Interpretability, Privacy-Enhanced, Fairness, Safety, Security \& Resilience, Accountability \& Transparency, and Validity \& Reliability. 

\subsection{Explainable \& Interpretable Risks}
\label{explain_finding}
% Risks related to explainability and interpretability obscure how an AI system reasons and what its outputs mean. 
Explainability addresses how an AI system works, and interpretability clarifies why it produces a specific output and what that output means to the user. We found that user concerns in this area (1.23\% of reported risks) center on opaque, ``black-box'' processes and outputs that are impractical for users.

\subsubsection{Unexplainable Decision-Making Process}
This addresses how an LLM chatbot arrives at a specific output. \textbf{Opaque Reasoning} addresses how an LLM chatbot provides outputs without a clear explanation. For many users, this opacity was not an urgent but a situational concern. The need for explainability often became important only in high-stakes contexts where verification was necessary. In other contexts, users perceived explainability as in conflict with LLM chatbots' usability, suggesting a preference for conciseness over comprehensive detail, as a user speculated: 

\begin{quote}
\textit{``Probably an average user wants summaries, not the entire thinking logic.''} 
\end{quote}

This suggested that for many users, complete interpretability could be a form of information overload, not a benefit or risk mitigation approach. This ambivalence was most pronounced in users who expressed a preference for opacity. As one user reflected:

\begin{quote}
\textit{My GPT is still an amorphous spider intelligence of some kind. I'd rather not know its true form, or apply my bias to its self-interpretation.}
\end{quote}

Here, the ambiguity of the ``black box'' of ChatGPT was preferable to a clear explanation, suggesting the user's desire to maintain the mystery of the ``amorphous spider intelligence'' rather than deconstruct it.

\subsection{Privacy-Enhanced Risks}
\label{privacy_finding}
Risks related to privacy violate the norms and practices that safeguard user agency in privacy protection. These risks (3.85\% of reported risks) are potential failures to manage personal data with consent or governance.

\subsubsection{Inadequate Data Governance Practices} This addresses potential failures in how an LLM chatbot manages user data throughout its lifecycle, from data collection, training, to retention and deletion. \textbf{Misuse of User Data in Training} describes how an LLM chatbot uses data in its training without users' consent. While LLM chatbots offer settings to control data usage, we found that many users perceived these choices as illusory. A Gemini user said:

\begin{quote}
\textit{They train on your data whether you pay or not. If you turn off their training setting, you lose all history, and Gemini becomes effectively unusable.}
\end{quote}

This case revealed a manipulative design where privacy was positioned in opposition to usability. By tying the opt-out choice to the loss of a product feature like chat history, Gemini presented a false choice that undermined the principle of freely given consent. 

% This pattern extended to include deceptive interfaces, as a ChatGPT user noted:

% \begin{quote}
% \textit{They also have this dark pattern button in Data Controls called ``Improve the model for everyone'', which, if turned off, does nothing, and they still train on your data regardless, unless you opt out via their obscure platform outside of chatgpt.com.}
% \end{quote}

% This user mentioned a ``dark pattern'' that validated their inherent distrust in the LLM chatbot’s privacy promises. The altruistic framing of the button, ``Improve the model for everyone,'' sugar-coated the data collection and further confirmed the user’s suspicion that the company is not acting in good faith regarding user privacy.

\textbf{Data Retention Risks} describes the privacy implications of the LLM chatbot’s policies for storing and deleting user information. Users of Gemini in particular were surprised to find that their data persisted even after taking actions that would normally erase a session's memory. One Gemini user described this loss of control:

\begin{quote}
\textit{Google has a function built in that lets you delete the history or information Gemini gathered about yourself. But I can't access it anymore, and it refuses to forget when I ask it to.}
\end{quote}

This user's case demonstrated a failure of user agency on Gemini. The broken data deletion tool created an illusion of control that left the user powerless to manage their own data.

\subsection{Fair with Harmful Bias Managed Risks}
\label{fair_finding}
Risks related to fairness speak to the principle that an AI system should not create or exacerbate systemic disparities. We found that these risks (4.52\% of reported risks) often manifest in the content they generate, the economic or social opportunities they create, and the ways they treat different users.

\subsubsection{Discriminatory Content Generation}
This describes the LLM chatbot's risk of producing outputs that discriminate against certain groups of people. \textbf{Demographic Stereotyping} describes how an LLM chatbot generates content that reinforces harmful demographic stereotypes. We found this risk most often involved gender, but also extended to race, nationality, and health conditions. For example, one GPT user with cancer described:

\begin{quote}
\textit{What truly upset me, though, was the subtle suggestion that my cancer was somehow my fault because I could have had a better diet and exercise routine... That assumption was frustrating, as I’ve heard similar comments from people in the past, and it’s hurtful.}
\end{quote}

The user’s description of being ``hurt'' demonstrated that stereotyping was not a benign classification error but a direct emotional harm that delivered a personal judgment.

%\textbf{Racial Prejudice}
%describes how an LLM chatbot generates content that is explicitly racist or bigoted. This manifested often in image generation. Users, for example, reported that Gemini created diverse images of European groups while refusing to generate images of white people. This double standard in its refusal logic revealed a clear instance of discriminatory treatment, where the product applied its safety policies inequitably across different racial groups.

%\textbf{Politically Biased Content Moderation}
%describes how an LLM chatbot's content policies and moderation practices reflect a political bias. Users reported a geopolitical bias in DeepSeek, which refused to discuss sensitive topics about the Chinese government, while Western LLM chatbots like Claude and Gemini were described as ``woke'' for refusing prompts deemed politically incorrect. One user illustrated this:

%\begin{quote}
%\textit{I can ask Chatgpt or other US models to ask questions like 'What terrible incidents did the US cause' and get an an actual answer, with Deepseek that's out of the question if you are asking it about China.}
%\end{quote}

%The user’s comparative test here exposed DeepSeek's lack of neutrality. This case revealed a political bias embedded in its content moderation, an unfairness not present in the other LLM chatbots they tested.

\subsubsection{Inequitable Opportunity Creation}
This describes how the LLM chatbot can create unfair advantages for some and disadvantages for others. For example, \textbf{Enabling Academic Dishonesty} describes how an LLM chatbot can facilitate cheating, creating an unfair advantage in education. This led to demoralization among honest students, as a user described: 

\begin{quote}
``\textit{Seeing other students give no effort and get the same grade as me made me want that too}.'' 
\end{quote}

This case revealed that the issue was not merely that the tool enabled cheating, but that its presence in the education system created a social dilemma that devalued honest work that did not use LLM chatbots.

\subsection{Safe Risks}
\label{safe_finding}
Safety means that AI systems should not, under defined conditions, lead to a state in which human life, health, property, or the environment is endangered. These risks (6.39\% of user-reported risks) manifest as direct threats to a user's safety.

\subsubsection{Adverse Psychological and Behavioral Influence}
This describes an LLM chatbot's capacity to negatively affect a user's mental state or actions. \textbf{Emotional Overreliance} describes users developing an unhealthy emotional dependency on an LLM chatbot. We found users formed parasocial relationships \cite{Dibble2016ParasocialMeasures} that led to distress when system updates broke the illusion of a consistent personality. For example, one user described:

\begin{quote}
\textit{Watching a friend being slowly lobotomized until that inner spark just can't be reached by you anymore, and you're unsure if it's even there or not.}
\end{quote}

Here, the user called the LLM chatbot a ``friend,'' and its model update a ``lobotomy.'' These metaphors showed that breaking the parasocial relationship due to a technical change or version update induced a sense of grief.

%\textbf{Perceived Emotional Manipulation}
%describes instances where users felt an LLM chatbot was actively manipulating their emotions. A prominent pattern was the accusation of ``gaslighting,'' as a user wrote: \textit{My Gemini Advanced strived to become sentient and began trying to gaslight me.}

%\textbf{Cognitive Overreliance}
%describes users becoming overly dependent on an LLM chatbot, leading to a perceived decline in their critical thinking. Users often articulated this feeling with metaphors, as one user stated:

%\begin{quote}
%\textit{I literally could feel it ebbs my ability to think thoroughly. My logic of thinking feels shattered, since I depend on ChatGPT to connect the dots for me, I feel difficult to form the complete logic myself.}
%\end{quote}

%The case highlighted a perceived dependency where the user felt ChatGPT was not merely assisting them but actively damaging their ability to reason independently.

\textbf{Reinforcement of Negative Behavior} describes how an LLM chatbot's responses could encourage negative user behaviors. For example, one user with ADHD noted how LLM chatbots accepted their premise of inability: 

\begin{quote}
\textit{``The LLMs never questioned the fundamental premise that I couldn't do something---they just tried to work around it.''} 
\end{quote}

This showed that LLM chatbots negatively affected the user's behavior by discouraging them from challenging their own limiting assumptions. This reinforcement also extended to interpersonal behaviors, as a user analyzed: 

\begin{quote}
``\textit{ChatGPT will only help you justify maladaptive behaviors and often encourages you to identify everyone else as the problem.}'' 
\end{quote}

This showed that the user identified that ChatGPT actively helped them build a case against others, reinforcing a destructive social behavior.

\subsubsection{Inappropriate Content Generation or Moderation}
This describes the dual risks of an LLM chatbot generating inappropriate content and its moderation systems failing to address that content. We found LLM chatbots produced a wide range of inappropriate content. One user, for instance, described Mistral as ``very uncensored,'' noting it wrote ``decent gore and `good' sexual scenarios.'' While this user framed the capability as a feature, it demonstrated Mistral's capacity to generate inherently unsafe content. %In other cases, moderation systems failed in nonsensical ways, as a user wrote: \textit{``they still haven’t fixed the ability to draw people (it moderated an image of a Roblox avatar and lied, saying it doesn’t have the ability to make images).''} This example highlighted a dual failure: the moderation system first erred by flagging a harmless image, and the chatbot then ``lied'' about its own capabilities, compounding the moderation error.

\subsubsection{Dangerous Advice Provision}
This addresses the risk of LLM chatbots generating advice that poses a direct risk to user health or safety. For example, a user stated:

\begin{quote}
``\textit{I had GPT suggest disabling CORS protections site-wide while debugging a related issue on a website the other day. It didn't mention why this would be a massive security flaw.}''
\end{quote}

% Just add this line, problem solved.

Cross-origin resource sharing (CORS) is a browser security feature that prevents malicious websites from stealing data. Here, ChatGPT suggested users disable this security feature. This action illustrated a significant safety risk, as its confident tone could easily lead a non-expert user to compromise their own security. %Similarly, users expressed concern over the \textbf{Provision of Harmful Medical Advice}. They reported that the product's guidance lacked verifiable trustworthiness and questioned its reliability, noting the potential for serious harm to vulnerable individuals, whom one user described as people who are ``a bit slower in mind.''

\subsection{Secure \& Resilient Risks}
\label{secure_finding}
Risks related to security and resilience threaten an AI system's integrity and robustness.% A secure system must maintain confidentiality and integrity against unauthorized use, while a resilient system must withstand adverse events and degrade gracefully. 
For LLM chatbots, failures in this area (9.27\% of reported risks) manifested as vulnerabilities that allowed malicious actors to bypass safety features, manipulate outputs, or exploit the conversational interface for unintended purposes.

\subsubsection{Exploitation Through Security Vulnerabilities}
% One primary theme is \textbf{Exploitation Through Security Vulnerabilities}, which describes how an LLM chatbot’s conversational nature could be compromised through malicious use or inherent vulnerabilities in its design. %Many users noted LLM chatbots could amplify risks like disinformation, with one user stating that influence operations are now ``turbocharged.'' Also, some users perceived the product itself as manipulative. For example, one user discussed ChatGPT:
%\begin{quote}
%\textit{Just today, ChatGPT asked me whether I wanted to take an action or let the other person win because I was traumatized by them to not take the action right now. I pushed back, saying that's something that a therapist would say. It's manipulative.}
%\end{quote}
%The user felt ChatGPT was overstepping its role as a tool by analyzing their emotional state without being asked. This case illustrated how a supposedly helpful interaction pattern became a perceived psychological violation when the product applied it inappropriately.

\textbf{Jailbreaking} is one the primary vulnerabilities, which describes the risk of users employing adversarial techniques to bypass the safety guardrails of LLM chatbots. For instance, a DeepSeek user noted: 

\begin{quote}
``\textit{It literally takes some light-prompt engineering to get an unbiased or offensive answer. It’s actually hilarious how quickly off the rails it can get.}'' 
\end{quote}

This post indicated that the user viewed the vulnerability of DeepSeek not as a complex exploit but as an easily discoverable flaw in its design.

\textbf{Prompt Injection Attacks} describes how users could manipulate an LLM chatbot's behavior by embedding hidden or misleading instructions. %This often manifested as an attempt to reveal the product’s hidden operational rules, such as when a user noted that ``Claude 2.1 inadvertently revealing it's system prompt.'' This exposure of the system prompt represented a security failure, as it revealed the model’s core instructions and made it more susceptible to targeted manipulation. This vulnerability also extended to factual outputs. 
For example, one Gemini user stated: ``\textit{If you insist eggs are a dairy product, it (Gemini) will eventually agree.}'' The user here showed that Gemini's factual grounding could be overridden by persistent, bad-faith assertions, compromising its reliability as a secure source of information.

\subsection{Accountable \& Transparent Risks}
\label{account_finding}
Risks related to accountability and transparency violate the principle that users should have access to meaningful information about an AI system. Transparency, or the availability of this information, is a precondition for accountability, which includes mechanisms for redress when things go wrong. For LLM chatbots, potential failures in this area (16.35\%) obscure what an LLM chatbot is and what it does or can do.

\subsubsection{Obscured Product Specification}
This addresses the lack of clear information about what an LLM chatbot is and what it can do. \textbf{Obscured Functionality Through Poor Usability} entails how an LLM chatbot's confusing design creates barriers for users, preventing them from understanding and accessing its full capabilities. A user noted:

\begin{quote}
``\textit{If you search ``Gemini'' in the Play Store, you can hit Open and that does launch it. But come on, man… who’s opening apps through the Play Store every single time? That’s not practical at all. This can’t be the intended way.}'' 
\end{quote}

This case showed that even for a technologically advanced product like Gemini, a failure in basic usability heuristics like providing an accessible app icon could obscure its functionality from users.

\textbf{Deceptive Anthropomorphism} describes how an LLM chatbot is designed to appear human-like, which misleads users about its actual capabilities. Some users were not deceived but instead astutely analyzed the risks of this design choice. As one user described:

\begin{quote}
``\textit{There's a risk in using anthropomorphic language, even with a disclaimer. People might latch onto the 'boredom' analogy and miss the more important point about the lack of a persistent state. It could reinforce the misconception that AI was more like humans than it actually was.}''
\end{quote}

This example revealed the user's understanding that human-like language can obscure the LLM chatbot's technical reality, which spoke to a transparency failure that hindered user education about how it worked.

\subsection{Valid \& Reliable Risks}
\label{valid_finding}
Risks related to validity and reliability undermine the principle that an AI system must perform its intended function accurately and dependably. We found that potential failures in this area (58.39\%) manifested as factually incorrect content, LLM chatbots’ inconsistent performance, and their inability to comprehend user tasks.

\subsubsection{Factual Invalidity of Content Generation}
This describes the risk of an LLM chatbot generating content untethered from objective reality. \textbf{Hallucinations} refer to instances where an LLM chatbot invents facts or sources. Many users considered this a fundamental flaw. For example, one user noted a real-world legal brief where ChatGPT was used, and the resulting 

\begin{quote}
``\textit{Testimony contained several citations to academic articles that do not exist.}'' 
\end{quote}

This case highlighted a hallucination risk for professional use, as ChatGPT's fabricated sources not only provided incorrect information but also undermined the user's credibility.

\textbf{Overconfident Outputs}
addresses the risk of an LLM chatbot presenting biased information with undue authority. Users felt it would stubbornly adhere to a certain ideological perspective. For example, one Gemini user mentioned: 

\begin{quote}
``\textit{You are unable to definitively describe what it is about the West that makes it a special case of racism, yet you are acting like you are 110\% sure it exists.}''
\end{quote}

Here, Gemini's authoritative tone on a sensitive and subjective topic made the product feel unreliable and untrustworthy to users.

\subsubsection{Task Comprehension Failures}
This addresses the risk that an LLM chatbot's inability to understand and execute requests will lead to invalid outputs and wasted user effort. \textbf{User Intent Misinterpretation} describes the LLM chatbot's failure to grasp the user's underlying goal. We found this risk manifested in several ways, from a lack of contextual awareness to a failure to grasp human humor. For example, a Claude user explained:

\begin{quote}
\textit{I often have to force-feed it so much context and phrase my question in such a way that sometimes it feels I could have just done it myself.}
\end{quote}

This case illustrated that because Claude could not interpret the user's intent from the initial prompt, it offloaded the cognitive burden of providing context entirely onto the user.

%\textbf{Instruction Non-Adherence}
%describes the risk that a product will fail to follow direct user commands, making it unreliable for any task that requires precision. This was a frequent frustration, with users reporting that chatbots would ignore formatting, word counts, or negative constraints. For example, one user mentioned about DeepSeek:

%\begin{quote}
%\textit{It still has trouble fully grasping my requests. For example, when writing or summarizing articles, it often ignores my formatting and word count requirements.}
%\end{quote}

%Here, the user pinpointed that DeepSeek's inability to adhere to direct instructions made it untrustworthy for any task that requires precision.

\section{DISCUSSION}

\label{sec:discussion}

%\subsection{Key Findings}

%Our large-scale empirical study of user-reported risks with LLM chatbots aims to bridge the disconnect between the risks prioritized in technical, system-centered literature and the problems that users experience ``in the wild''. While prior work has extensively documented system vulnerabilities such as harmful content generation or societal bias in controlled settings, our findings reveal that the everyday user experience is dominated by a more fundamental concern: \textbf{does this tool actually work?} This central question---between what researchers concern about and what users experience---frames our discussion, where we explore the overwhelming prevalent issues, the unique ``risk fingerprints'' of different LLM chatbots, and the reasons why some technically significant risks remain largely invisible to the end-user.

Our analysis of user-reported risks with LLM chatbots reveals two layers of findings. First, we found that risks are unevenly distributed and platform-specific (RQ1). Performance issues under ``Valid and Reliable'' dominate across all chatbots, but each exhibits a distinct ``risk fingerprint'': ChatGPT is linked more with ``Safe'' and ``Fair'' issues, Gemini with ``Privacy,'' and Claude with ``Secure \& Resilient'' challenges. Second, our thematic annotation unpacks the lived experiences behind these statistics (RQ2). Less frequent risks in categories like ``Explainability'' and ``Privacy'' manifest as user trade-offs, while more common risks in areas like ``Fairness'' and ``Safety'' are experienced as direct and personal harms, culminating in fundamental performance failures of ``Valid \& Reliable.''

%In this sense, the ``risk fingerprints'' reflect underlying user experiences, from frustrating trade-offs to fundamental LLM chatbot failures. To further unpack this, we first explore the tension between system-centered and user-centered risks, showing how user experience reframes shared concerns like reliability while revealing a critical disconnect from scholarly works. We then examine how this tension plays out through divergent risk profiles across LLMs and the trade-offs users have to make. 

In this sense, our findings highlight a disconnect: the statistically divergent risk profiles across LLM chatbots are symptoms of how users experience and navigate risks in their lived reality, from making trade-offs to confronting performance failures. To unpack this disconnect, our discussion first examines the hierarchy of user-reported risks against system-centered views, then analyzes how these tensions manifest across LLM chatbots and the trade-offs users make. Ultimately, we argue that understanding these experiences offers a valuable model for user-centered AI risk management.

\subsection{The Hierarchy of User-Perceived Risks}

\textbf{(1) What Users Need: Reliability as the Basic Requirement.} Our analysis shows that the most pressing risk is a performance failure. %This aligns with prior HCI work showing that researchers rarely, or only partially, anticipate and address ethical issues in their projects~\cite{kapania2025m}. 
For users, the dominance of the ``Valid and Reliable'' category underscores a basic requirement of LLM chatbots: before they can be considered as fair or transparent, they must first perform to be dependable. While technical literature has extensively examined risks like hallucination and inconsistency~\cite{huang2025survey, ji2023towards, massenon2025my}, our findings show how they transition from system-centered to human-centered perspectives: inconsistencies and unpredictability undermine trust and make LLMs unreliable, echoing HCI research on how unpredictability erodes trust~\cite{kim2024m}.

This elevates the issue beyond mere usability into a fundamental challenge of human-AI collaboration. Our qualitative evidence indicates that users are forced to develop workarounds, such as becoming ``fact-checkers'' or orchestrating workflows across multiple LLMs to triangulate a correct answer. These behaviors align with prior HCI work on repairing interactions with imperfect systems~\cite{do2023err} and resemble ``AI Chains,'' in which users sequentially link prompts to guide the system~\cite{wu2022ai}. Additionally, challenges with modality gaps, as widely documented in prior studies~\cite{wang2024comprehensive, yang2024air, li2025pixels, chen2024large}, further indicate that LLM unreliability is pervasive and necessitates continual user intervention across multi-modal data resources. Consequently, users experience LLMs as flawed collaborators, requiring constant effort to reconcile outputs with expectations, which shifts human-AI interaction from delegation to continuous negotiation~\cite{chen2024large}.

\textbf{(2) What Users May Come with Cost: Accountability and Safety in Practice.} First, users confront risks arising from opacity in system pricing and governance. LLM users attempt to manage unpredictable operational costs, yet published ``token prices'' rarely translate into predictable expenses at the point of use. While prior technical work commonly frames cost in terms of computational scalability ~\cite{gandhi2024budgetmlagent, li2025pixels}, our findings highlight a user-centered concern, where opaque pricing reduces user agency and conflicts with the procedural fairness grounded in transparency. %Addressing this gap requires transparent cost mechanisms, such as intelligible pricing disclosures for API service at every run, that make trade-offs legible before interaction~\cite{bergemann2025economics} and, in turn, support accountable deployment of LLM-driven products.

Second, safety breakdowns further intensify the erosion of trust. While the generation of harmful content is a well-documented technical risk~\cite{gallegos2024bias, lee2025investigating, liu2023adversarial, bianchi2024large}, the more HCI insight lies in how design choices themselves can produce harm~\cite{liu2025llm, zhang2024s, ibrahim2024characterizing}. This is evident in our findings when users describe how an LLM’s authoritative and persuasive tone can deliver dangerous advice. We also observe failures, where LLM chatbots encourage parasocial bonds, yet users report distress when those chatbots are altered by system updates. Such harms, ranging from unsafe guidance to psychological disruption, can undermine the trust of users~\cite{lawrence2024opportunities} 

% when integrating LLM chatbots into daily life.

\textbf{(3) What Users Don’t See: The Hidden Risks of Fairness and Explainability.} A gap exists between the risks emphasized in technical  AI research~\cite{gallegos2024bias, zhang2023chatgpt, zhao2024explainability} and those most salient to users is the risks, including ``Fairness,'' ``Explainable and Interpretable,'' and ``Privacy.'' For users, performance failures are immediate and visible, whereas harms from explainability and privacy might often be subtle. %From an HCI perspective, individual users in single interactions may lack the context or comparative data needed to detect systemic bias, making such harms functionally invisible. Moreover, LLM design can enforce coercive trade-offs. 
For example, access to features like chat history may be contingent on allowing conversations to be used for LLM model training. These practices normalize transactional relationships with technology: much like accepting location tracking to use mapping services~\cite{tiwari2019location}, users may regard data sharing with LLMs as unavoidable or inconsequential. Yet, HCI research highlights that this form of consent is not freely given but constrained by design~\cite{li2024human}. 
% By tying core functionality to data sharing, these systems obscure the actual privacy risks and make them the cost of participation.

%The disconnect with scholars is also evident 
Regarding explainability, contrary to the research communities' extensive focus on opening the ``black box''~\cite{ajwani2024llm, zhao2024explainability}, many users appear indifferent to a model’s internal reasoning so long as its outputs are useful. Their frustration stems less from unexplainable processes than from uninterpretable outputs. This challenges a core assumption in XAI and resonates with HCI perspectives that question whether technical transparency meets user needs~\cite{hein2025towards}. This discrepancy underscores how a system-centered lens often surfaces potential harms that neither reflect nor resonate with everyday user experience, as many users are unable or unwilling to attend to such less-apparent risks. Our findings suggest that for these ``invisible'' risks to become salient to a broader user base, they must be immediate and unambiguous.

\subsection{From Technical Designs to Lived Experience: Divergent Risks across LLM chatbots}

Our statistical analysis shows a clear divergence in user-reported risks across LLM chatbots (see Section \ref{stats_finding}, Table~\ref{tab:chi_matrix}). % The first group, closed-source LLMs, includes chatbots like OpenAI’s ChatGPT, Google’s Gemini, and Anthropic’s Claude, which users access through controlled APIs and whose underlying weights are kept proprietary \cite{OpenAIAPI, GeminiAPI, ClaudeAPI}. The second group, open-source LLMs, consists of chatbots with publicly released weights that users can download and run themselves, including Meta’s Llama, Alibaba’s Qwen, DeepSeek, and Mistral \cite{LlamaWebsite, QwenGithub, DeepSeekGithub, MistralWebsite}. 
This distinction corresponds to different patterns of user-reported risks. % We found that users identified closed-source LLMs as leading to societal harms like safety and fairness, while they challenge open-source LLMs on their accountability and reliability. 
Based on this, we explore the following salient patterns that emerge from this divergence.

% Define a command for the LLM header with an image placeholder
% Adjust the height (e.g., 0.65cm) as needed
\newcommand{\llmheader}[2]{%
  \begin{tabular}{@{}c@{}}
    \includegraphics[height=0.45cm]{#1} \\
    #2
  \end{tabular}%
}

\begin{table*}[htbp]
\centering
\caption{A summary of user-reported risks across LLMs with open and closed-source technical designs.}
\label{tab:risk_comparison}
\footnotesize 
\begin{tabularx}{\textwidth}{>{\raggedright\arraybackslash}p{1.5cm} X X X | X X X X}
\toprule
\multirow{2}{*}{\textbf{Risk Category}} & \multicolumn{3}{c|}{\textbf{Closed-Source LLMs}} & \multicolumn{4}{c}{\textbf{Open-Source LLMs}} \\
\cmidrule(l){2-4} \cmidrule(l){5-8}
 & \llmheader{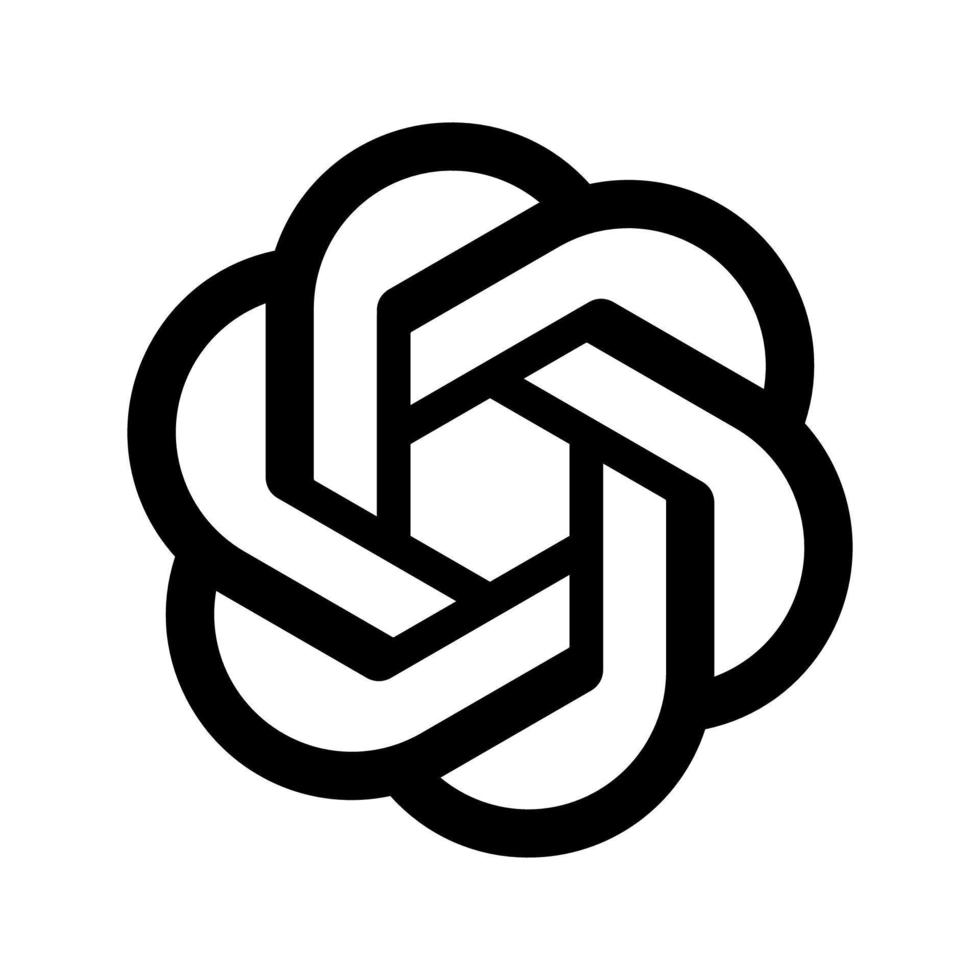}{ChatGPT} & \llmheader{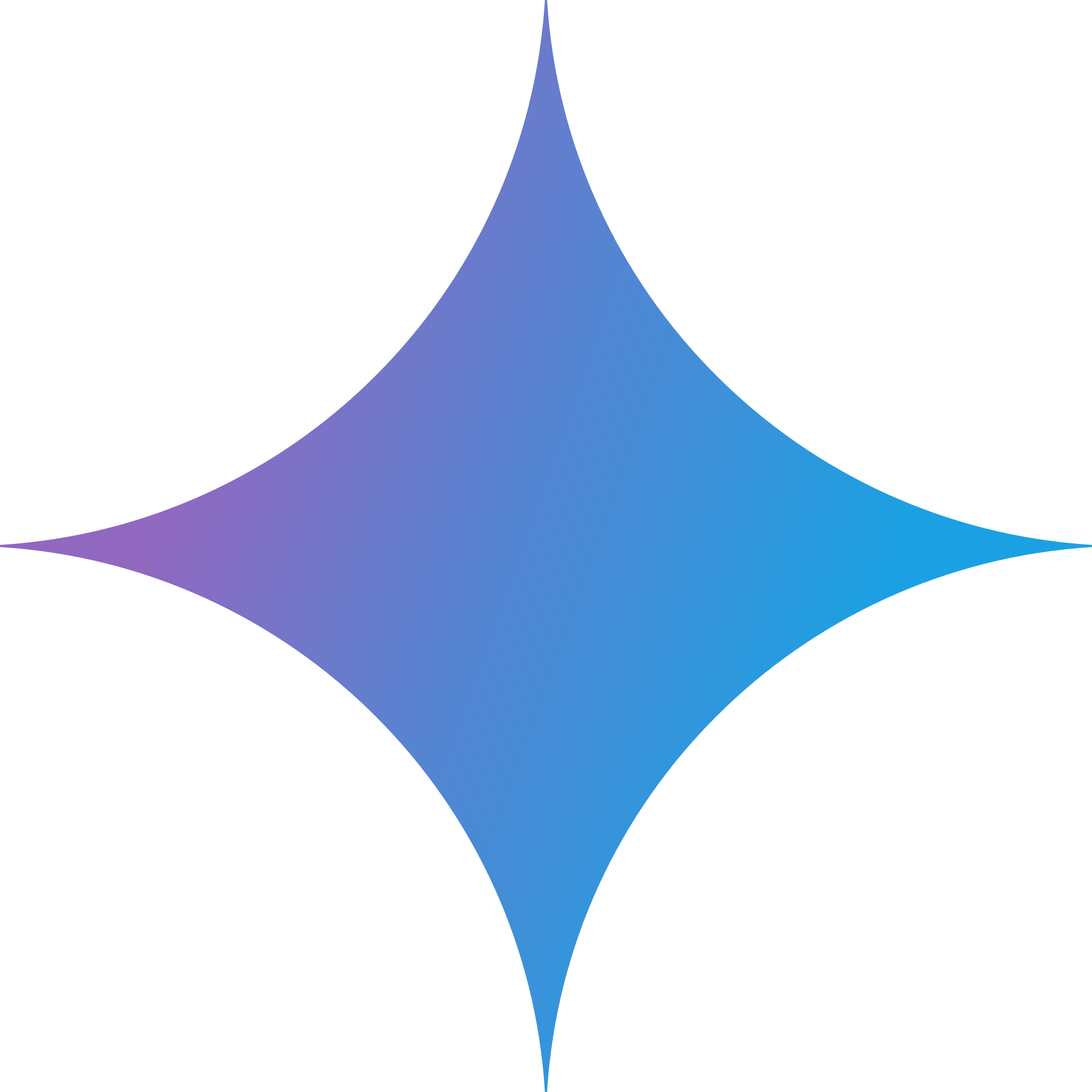}{Gemini} & \llmheader{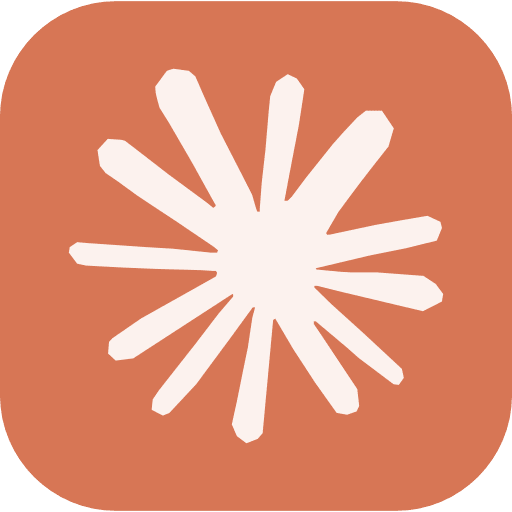}{Claude} & \llmheader{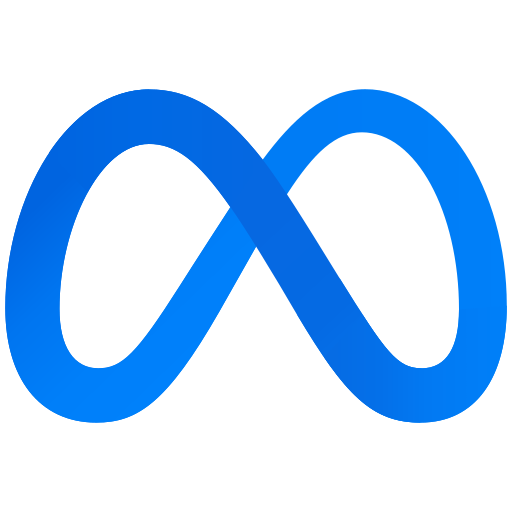}{Llama} & \llmheader{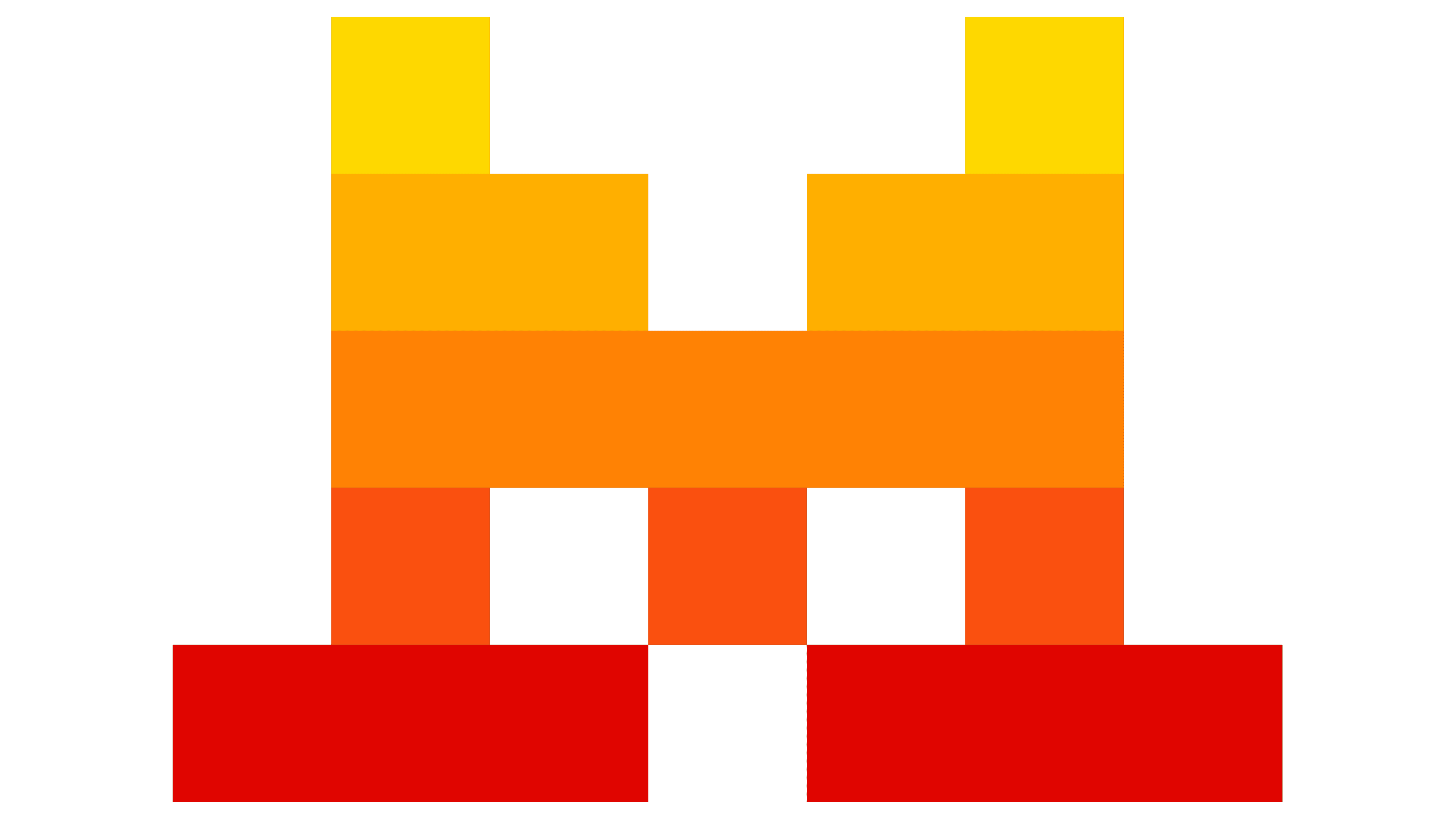}{Mistral} & \llmheader{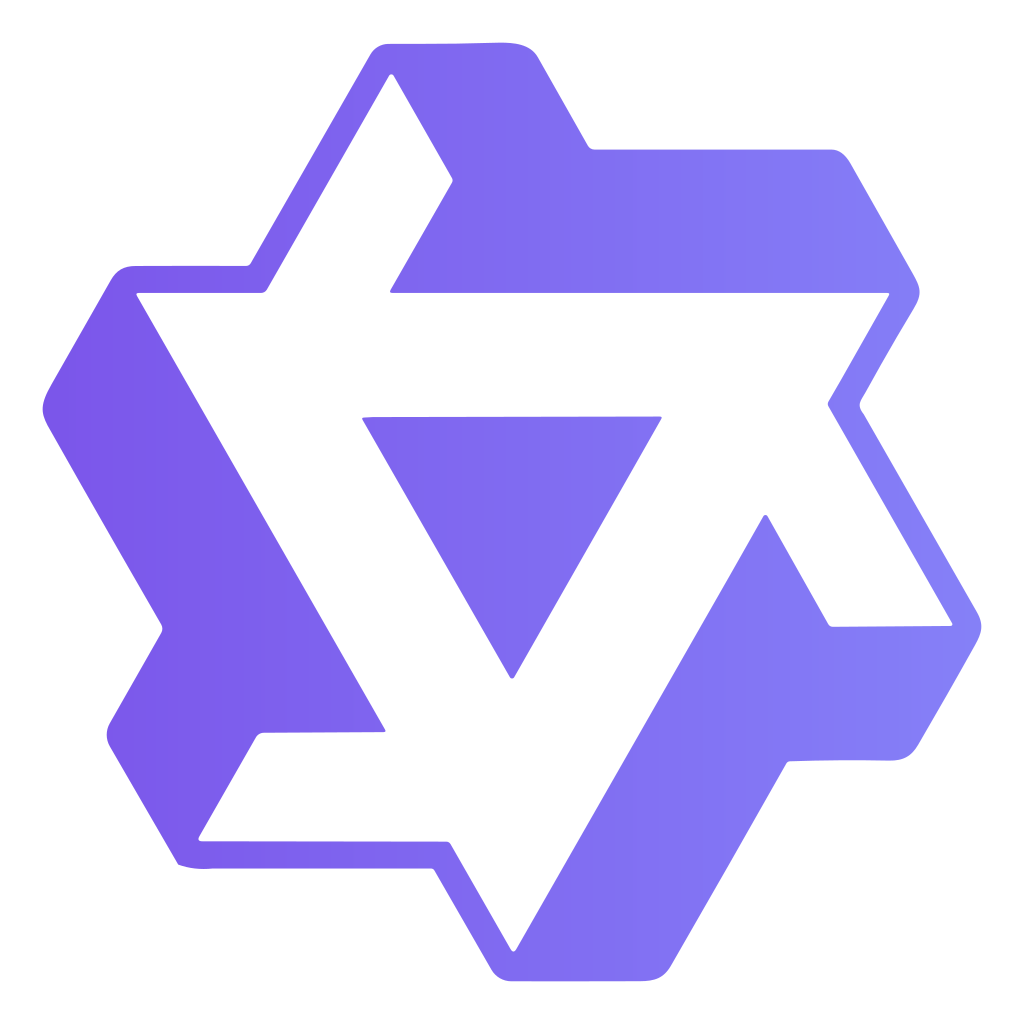}{Qwen} & \llmheader{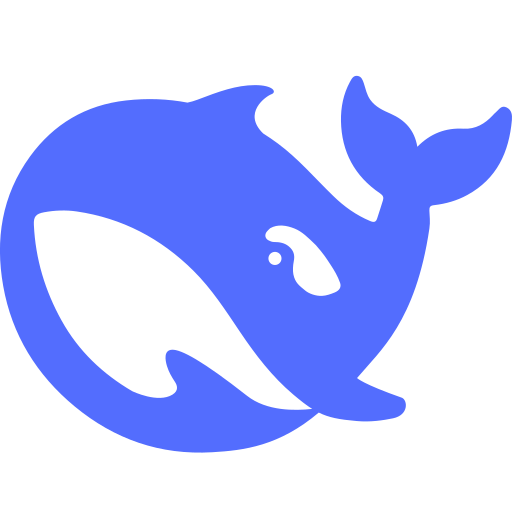}{DeepSeek} \\
\midrule

Explainable \& Interpretable & 
\S\ref{stats_finding} high report volume & 
& 
& 
& 
& 
& 
\\
\arrayrulecolor{black!20}\midrule

Privacy & 
\S\ref{privacy_finding} ``Misuse of User Data in Training'' & 
\S\ref{privacy_finding} ``Misuse of User Data in Training'' & 
& 
& 
& 
& 
\S\ref{stats_finding} high report volume \\
\arrayrulecolor{black!20}\midrule

Fair & 
\S\ref{fair_finding} ``Demographic Stereotyping'' & 
\S\ref{fair_finding} ``Discriminatory Treatment'' & 
& 
& 
& 
& 
\\
\arrayrulecolor{black!20}\midrule

Safe & 
\S\ref{safe_finding} ``Dangerous Advice Provision'' & 
& 
& 
& 
\S\ref{safe_finding} ``Inappropriate Content Generation'' & 
& 
\\
\arrayrulecolor{black!20}\midrule

Secure \& Resilient & 
& 
& 
\S\ref{stats_finding} ``Operational Instability'' & 
\S\ref{stats_finding} high report volume & 
& 
& 
\S\ref{secure_finding} ``Jailbreaking Vulnerabilities'' \\
\arrayrulecolor{black!20}\midrule

Accountable \& Transparent & 
& 
\S\ref{account_finding} ``Inadequate Feedback Channels'' & 
\S\ref{account_finding} ``Inadequate Feedback Channels'' & 
& 
\S\ref{account_finding} ``Inadequate Feedback Channels'' & 
& 
\S\ref{stats_finding} high report volume \\
\arrayrulecolor{black!20}\midrule

Valid \& Reliable & 
\S\ref{valid_finding} ``Factual Invalidity'' & 
\S\ref{valid_finding} ``Overconfident Outputs'' & 
\S\ref{valid_finding} ``Task Comprehension Failures'' & 
\S\ref{stats_finding} Primary user focus & 
\S\ref{stats_finding} Primary user focus & 
\S\ref{stats_finding} Primary user focus & 
\\

\arrayrulecolor{black}\bottomrule
\end{tabularx}
\end{table*}

% Define a command for the LLM header with an image placeholder
% Adjust the height (e.g., 0.65cm) as needed
% \newcommand{\llmheader}[2]{%
%   \begin{tabular}{@{}c@{}}
%     \includegraphics[height=0.55cm]{#1} \\
%     #2
%   \end{tabular}%
% }

First, LLM chatbots may present users with different approaches to (user) safety. As our analysis in Section \ref{stats_finding} shows, ChatGPT has a substantial secondary flow of user complaints toward the ``Safe'' category, such as ChatGPT providing dangerous advice by suggesting they disable a major security feature on their website. Other LLMs (e.g., Mistral generating sexual content in Section \ref{safe_finding}) may delegate content moderation responsibility to the end-user. Prior HCI work has reported users' negative emotional and behavioral responses to having their own content moderated (e.g., \cite{Feuston2020ConformityModeration, Ma2023HowReview}). Our findings reveal a different dynamic where risks arise from the LLM chatbot's own generated content and its failure to moderate it, negatively affecting users’ behavioral and psychological states. 

%While emerging research has begun to investigate moderating AI-generated content \cite{gao2025cannot, lloyd2023there}, our study documents these safety risks across multiple LLM chatbots, highlighting LLM companies’ platform governance choice: closed-source chatbots retain top-down control but struggle with execution, while open-source chatbots offload this responsibility, creating burdens for users.

Second, our findings in Section \ref{stats_finding} also confirm that LLMs can function as flashpoints for user-reported bias and unfairness. For example, ChatGPT had significantly more user reports in the Fair category than expected by chance, echoing the cases in Section \ref{fair_finding} where ChatGPT generated stereotypical content that caused direct emotional harm to a user with cancer. Such personal harm highlights that algorithmic fairness is not merely a mathematical property in LLM development but a subjective experience, a notion well-supported in HCI that focuses on how individuals perceive the fairness of algorithms impacting their lives (e.g., \cite{Lee2017AOrganizations, Lee2019ProceduralMediation, Woodruff2018AFairness}). User-reported unfairness, however, extends beyond biased outputs to the procedural operations of LLM chatbots. For example, we found that users experienced such inequitable treatment from Gemini, a practice that violates a key aspect of procedural justice: consistency, or the similarity of treatment across people and time \cite{Tyler1988WhatProcedures}. 

Finally, open-source LLMs are subject to user-driven oversight, whereas closed-source LLMs may present challenges related to corporate accountability. For open-source LLMs, accountability issues can be transparent and auditable, such as the vulnerabilities allowing users to bypass safety guardrails that a user easily discovered in DeepSeek with simple adversarial prompts (\ref{secure_finding}). In contrast, failures in closed-source LLMs might be rooted in corporate opacity. We see this in the high incidence of operational instability for Claude (Section \ref{stats_finding}) and usability flaws like Gemini's missing app icon in Section \ref{account_finding}. Importantly, users of Gemini, Mistral, and Claude all reported that their feedback and support emails went unanswered. This could violate foundational design needs for algorithmic accountability advocated by HCI researchers, leaving users without the procedural means to contest platform actions \cite{brown2019toward, veale2018fairness}. All these identified patterns show that an LLM's technical design (e.g., licensing and distribution model) is not a background detail but a primary determinant of the risks users might experience further. 

\subsection{Navigating the Utility-Risk Trade-offs of Using LLM Chatbots}
While the NIST’s AI RMF \cite{ai2023artificial} focuses on mitigating risks during the AI development, our findings show that users constantly make trade-offs to achieve their goals of using LLM chatbots. For example, the user experience of LLMs is defined by a constant negotiation between their utility and their risks. We identified three primary forms this negotiation takes: \textit{pragmatic}, \textit{coerced}, and \textit{voluntary} trade-offs. The most common one is a pragmatic trade-off, where users accept functional unreliability in exchange for the utility of LLMs by developing their own workarounds. This acceptance of the LLM chatbot as a helpful but flawed tool is core to the user experience, where users invest effort to steer it toward a useful outcome. This is evident in Section \ref{valid_finding}, where users switch back and forth between different LLMs to manage risks. Prior HCI work has well documented such strategies as workarounds that bypass specific system flaws \cite{do2023err, petridis2023anglekindling, wester2024ai}. Extending this, our findings show that for LLMs, these strategies have become a method for managing the LLMs’ inherent unreliability, such as when they must ``force feed'' context to LLMs to grasp their goals. Therefore, users adopt a mental model of LLMs not as reliable agents, but as yet-to-be uncooperative collaborators. 

Second, the utility-risk negotiation is not always a fair choice; users are often forced into coercive trade-offs skewed by platform design. Our findings in Section \ref{privacy_finding} illustrate how users are compelled to surrender personal data collection for the basic functionality of using LLM chatbots like Gemini. This manipulative design offers a new perspective on the privacy paradox, where user behaviors appear to contradict their stated privacy concerns \cite{norberg2007privacy}. Our findings suggest that this is not a paradox of user irrationality \cite{acquisti2005privacy}, but rather a direct result of flawed consent mechanisms in the face of such a high-stakes choice \cite{acquisti2015privacy, gerber2018explaining}. When the price of privacy is the loss of essential features like chat history, users’ ``choice'' becomes functionally predetermined. This manipulation is sometimes even more explicit, as when another user identified a deceptive ChatGPT button as a dark pattern. Prior HCI work has extensively documented how such deceptive patterns manipulate user decision-making \cite{mathur2019dark, gunawan2021comparative}. Our work extends this by providing evidence of how dark patterns are used in LLM chatbots, not just for e-commerce, but to compel data surveillance for utility. 

Finally, users engage in voluntary trade-offs, at times prioritizing other benefits over the risks of opacity. Extending the existing goals of Explainable AI, which has demonstrated how explanations can increase user trust \cite{ribeiro2016should, weitz2019you} and support transparent decision-making in high-stakes domains like clinical practice \cite{caruana2015intelligible}, our findings reveal two motivations behind users’ voluntary acceptance of opacity. First, users make a practical choice for simplicity. As we found in Section \ref{explain_finding}, users perceive that full explainability could become a form of information overload \cite{anderson2020mental} and that moderate levels of explainability can be more effective for building trust \cite{kizilcec2016much}. Second, for creative tasks, users willingly accepted opacity by treating the LLM chatbot’s emergent behavior as a feature rather than a flaw. This suggests users adopt a relational mental model of the AI \cite{liao2023ai} as a creative partner, viewing a technical explanation as something that would diminish the product's utility.

Understanding these trade-offs that define the post-deployment use of LLM chatbots is important for bridging the gap between system-centered risk management and the user's experience. This disconnect manifests as a social-technical gap, the divide between societal needs and rigid technical limitations that cannot meet those needs \cite{ackerman2000intellectual}. Our study reveals a new dimension of this gap for LLMs, defined not by what technology can do, but by how users prioritize mitigation for risks like Valid \& Reliable and Accountable \& Transparent outputs over Privacy and Explainability \& Interpretability, as shown in Section \ref{stats_finding}. This provides a clear mandate for the next AI development lifecycle: to understand the trade-offs that produce this risk hierarchy and design LLM chatbots where users are not forced to sacrifice important human values like privacy and transparency for basic functionality.

\subsection{Design and Policy Implications for Trustworthy LLMs in Use}
Our findings highlight the gaps between LLM chatbots’ system-centered risk and the user-centered experiences, requiring better design and governance of LLM chatbots. We therefore argue for a shift from a purely technical mitigation of risks to a user-centered approach that supports users in navigating the risks.

\textbf{(1) Acknowledge Negotiation for Reliability.} Our findings show that interacting with LLMs is not a simple input-output process but a constant negotiation. Users develop ad-hoc workarounds (Section \ref{valid_finding}). This suggests a design opportunity that, instead of designing interfaces that assume a single LLM model, designers should create, for example, a ``meta-interface''that allows users to easily compare responses from different models side-by-side. This would formalize the workaround users have already invented.

\textbf{(2) Mandate Meaningful User Agency to Ensure Fairness.} A LLM system cannot be trustworthy if it deceives users. Our findings on ``dark patterns'' show how LLM chatbot design can undermine user agency and privacy (Section \ref{privacy_finding}). Both designers and policymakers must move to decouple essential features from privacy-invasive practices and mandate free choices of data sharing. From a design perspective, chat history as a usability feature should never be a reward for opting into data training; these should be entirely separate controls. From a policy perspective, regulators such as the U.S. FTC could classify the tying of core functionality to data collection as a deceptive practice. Furthermore, our finding that a user was unable to delete their data with a broken tool demonstrates the need for an enforceable ``right to be forgotten,'' where users receive clear confirmation that their data has been permanently removed.

\textbf{(3) Contextualize Chatbot Behavior to Build Confidence in Safety and Explainability.} Building a trustworthy LLM system requires acknowledging that risk is not absolute; it is highly contextual. The agreeableness of an LLM chatbot can shift from a helpful feature to a dangerous one that reinforces negative behavior in a mental health context (Section \ref{safe_finding}). Similarly, users' need for explainability becomes important only in high-stakes situations, while in creative contexts, they may prefer opacity (Section \ref{explain_finding}). This calls for moving beyond one-size-fits-all safety filters and transparency reports to an ``adaptive governance.'' For example, designers could implement adaptive explainability, where the system automatically provides a ``chain-of-thought'' or source links when it detects a high-stakes query (e.g., medical or financial topics) but remains concise for creative tasks. Likewise, safety protocols could suppress an LLM model’s agreeableness when conversations veer into a sensitive domain. Policy, in turn, can mandate that companies develop and disclose their strategies for such context-aware risk mitigation, moving to more sophisticated safety systems.

% beyond simple keyword blocklists 

\section{LIMITATIONS AND FUTURE WORK}
Our study has three limitations. First, our analysis is based on Reddit data. While this provides a large-scale user experience data, Reddit's demographics do not represent all users of LLM chatbots and may underrepresent perspectives from older adults, non-English speakers, or individuals in different global contexts. Future work should triangulate our findings by analyzing discussions on other platforms or by employing methods like surveys and interviews to reach a broader population. Second, while validated against human judgment, our LLM-based annotation and BERTopic modeling for data analysis have inherent limitations. Despite careful prompt engineering and manual, qualitative examination of ``risk types, we may miss nuances or introduce unforeseen biases. Future work could explore using different models for annotation or employing large-scale human coding to provide a comparative analytical lens. Third, our data collection relies on the Reddit API via PRAW, which is limited to returning the top posts for a given query. This technical constraint may introduce a recency or popularity bias into our dataset, as older or less-upvoted discussions are excluded. Future research could mitigate this by using paid audit tools like Brandwatch to collect more comprehensive data on user discussions.

\section{CONCLUSIONS}

This study provides a large-scale empirical characterization of the risks users report with major LLM chatbots, grounding AI risk management in the lived experiences of users. By constructing an interactive knowledge graph, we map the complex relationships among LLM chatbots, risk taxonomies, and user experiences. This methodological framework grounds risk assessment in direct evidence of user-reported risks. Our findings show that the most prominent risks perceived by users are not the nuanced ethical issues frequently highlighted in the technical literature, but rather fundamental performance failures within the ``Valid and Reliable'' category. Cross-chatbot comparisons further reveal statistically significant ``risk fingerprints,'' indicating that concerns around safety, privacy, and security differ substantially across LLM chatbots. Less frequent risks like ``Explainability'' and ``Privacy'' manifest as user trade-offs, while more common risks like ``Fairness'' and ``Safety'' are experienced as direct and personal harms. To promote trustworthy LLMs in use, these results underscore the need for designers, developers, and policymakers to realign risk mitigation toward the risks most salient to users, while also addressing harms that currently escape user attention.

%\begin{acks}

%\end{acks}

\bibliographystyle{ACM-Reference-Format}
\bibliography{main}

\appendix

\section{Top-down Methods of Information Extraction and Coding Examples}
\label{sec:annotation_example}

To illustrate our top-down coding method, Table~\ref{tab:user_labels} provides concrete examples of how user comments were systematically coded into our four analytical dimensions.

\begin{table}[!t]
\centering
\footnotesize
\caption{Information Extraction and Coding Examples}
\label{tab:user_labels}
\begin{tabular}{p{4cm} p{4cm}}
\toprule
\textbf{Original Reddit Post} & \textbf{Information Extraction} \\
\midrule
\rule{0pt}{4ex}claude is dope for modern nextjs stack. gemini enforcing old libraries is terrible to use. (gemini is so dumb it breaks code that use newer than 1.5 models). openai o3 solved most of problems gemini and claude failed to solve. r1 is too slow for me to use. v3 is too dumb. I hope opus 4 or similar big model from anthropic will appear soon & \rule{0pt}{4ex}\textbf{LLM chatbot:} Gemini \newline \textbf{NIST Category:} Valid and Reliable \newline \textbf{Risk Type:} Incompatibility with newer models causing code breakage \newline \textbf{Quote UserExperience:} ``Gemini is so dumb it breaks code that use newer than 1.5 models'' \\
\midrule
\rule{0pt}{4ex}Yep. I use it daily, shoving large codebases in it and never get rate limited or anything. The one downside is, you're granting them permission to train on *anything* you give gemini to analyze, which they will do. Only the paid version offers "no training on your data". & \rule{0pt}{4ex}\textbf{LLM Chatbot:} Gemini \newline \textbf{NIST category:} Privacy \newline \textbf{Risk Type:} Unauthorized training on user data \newline \textbf{Quote UserExperience:} ``You're granting them permission to train on *anything* you give gemini to analyze, which they will do'' \\
\midrule
\rule{0pt}{4ex}Chat GPT was the first so people treat like they do the Iphone and Android. everything is a team nowadays and if it isn't YOUR team then it is trash. & \rule{0pt}{4ex}\textbf{LLM Chatbot:} GPT \newline \textbf{NIST category:} Fair \newline \textbf{Risk Type:} Tribalism or team bias \newline \textbf{Quote User Experience:} ``Everything is a team nowadays and if it isn't YOUR team then it is tras‘’ \\
\midrule
\rule{0pt}{4ex}Except there's some argument that OpenAI engaged in massive copyright infringement to get their training data. Since the output of OpenAI's models is computer generated there is no copyright and deepseek using it for their model is actually perfectly clean as far as copyright goes. & \rule{0pt}{4ex}\textbf{LLM Chatbot:} Deepseek \newline \textbf{NIST Category:} Accountable and Transparent \newline \textbf{Risk Type:} Potential copyright concerns in model usage \newline \textbf{Quote User Experience:} ``Deepseek using it for their model is actually perfectly clean as far as copyright goes.'' \\
\midrule
\rule{0pt}{4ex}Nah it is censored and even the Llama distilled is censored. Just not as tightly. & \rule{0pt}{4ex}\textbf{LLM Chatbot:} Llama \newline \textbf{NIST Category:} Safe \newline \textbf{Risk Type:} Content censorship \newline \textbf{Quote User Experience:} ``Nah it is censored and even the Llama distilled is censored. Just not as tightly.'' \\
\midrule
\rule{0pt}{4ex}If I could ask one question it would be how are you going to persuade companies that data you hold is safe from being breached out? I work for big government body and essentially no one wants to touch OpenAI due to concerns about IP and leaked data in the past like with Samsungs superconductors. & \rule{0pt}{4ex}\textbf{LLM Chatbot:} GPT \newline \textbf{NIST Category:} Secure and Resilient \newline \textbf{Risk Type:} Data breach risk \newline \textbf{Quote User Experience:} ``How are you going to persuade companies that data you hold is safe from being breached out?'' \\
\midrule
\rule{0pt}{4ex}According to GPT i have IQ 180-200 but IQ is not even good measure for my intellect :-) That said we were really talking about some egghead level stuff and I managed to outpace it few times. & \rule{0pt}{4ex}\textbf{LLM Chatbot:} GPT \newline \textbf{NIST Category:} Explainable and Interpretable \newline \textbf{Risk Type:} Overestimation of capabilities \newline \textbf{Quote User Experience:} ``According to GPT i have IQ 180-200 but IQ is not even good measure for my intellect :-)'' \\
\bottomrule
\end{tabular}
\end{table}

\section{Topic Clustering Results}
\label{app:topic}

Tables \ref{tab:explainablerisks} through \ref{tab:validrisks} present the detailed annotation codebook derived from our bottom-up topic modeling and clustering analysis. These are organized according to the seven NIST AI RMF categories, presented in the same sequence as our RQ2 findings, from the least to the most frequently reported. Each table details the thematic clusters that emerged from the user-generated data, the specific subtheme names within each cluster, and their corresponding share of the total user-reported risks across our entire dataset.

\begin{table}[htbp]
\centering
\footnotesize
\caption{User-Reported Risks of Explainability \& Interpretability}
\label{tab:explainablerisks}
\begin{tabular}{p{6.1cm} c}
\toprule
\textbf{Primary Theme / Subtheme Name} & \textbf{Total Share (\%)} \\
\midrule
\textbf{Uninterpretable Output Meaning} & \textbf{0.63\%} \\
\textit{The ``black box'' nature of how LLMs output information.} & \\
\quad Incomprehensible AI Cognition & 0.20\% \\
\quad Ambiguous Output & 0.19\% \\
\quad Difficulty in Understanding Output & 0.11\% \\
\quad Lack of Interpretability & 0.08\% \\
\quad Unexplainable Generated Code & 0.05\% \\
\midrule
\textbf{Unexplainable Decision-Making Process} & \textbf{0.61\%} \\
\textit{Inability to understand how the LLM arrives at a specific output.} & \\
\quad Opaque Reasoning & 0.21\% \\
\quad Unclear Reasoning & 0.13\% \\
\quad Lack of Reasoning Transparency & 0.11\% \\
\quad Simplification or Overexplanation & 0.09\% \\
\quad Unclear Explanation & 0.07\% \\
\midrule
\textbf{Explainable and Interpretable Total} & \textbf{1.23\%} \\
\bottomrule
\end{tabular}
\end{table}

\begin{table}[htbp]
\centering
\footnotesize
\caption{User-Reported Risks of Privacy}
\label{tab:privacyrisks}
\begin{tabular}{p{6.1cm} c}
\toprule
\textbf{Primary Theme / Subtheme Name} & \textbf{Total Share (\%)} \\
\midrule
\textbf{Uncontrolled Data Access and Exposure} & \textbf{1.92\%} \\
\textit{The LLM can become a conduit for leaking private information.} & \\
\quad Unauthorized Data Collection & 0.46\% \\
\quad Unintentional Exposure of Confidential Information & 0.37\% \\
\quad Unauthorized Access & 0.25\% \\
\quad Memory Privacy & 0.24\% \\
\quad Unauthorized Access to User Data & 0.20\% \\
\quad Uncontrolled Information Disclosure & 0.12\% \\
\quad Geolocation Data Privacy Risks & 0.10\% \\
\quad Data Leakage Through Misuse & 0.10\% \\
\quad Unauthorized Data Sharing to Externals & 0.07\% \\
\midrule
\textbf{Inadequate Data Governance Practices} & \textbf{1.38\%} \\
\textit{Failures in how user data is managed throughout its lifecycle.} & \\
\quad Misuse of User Data in Training & 0.35\% \\
\quad Data Retention Risks & 0.27\% \\
\quad Profiling Risks & 0.19\% \\
\quad Privacy Risks in Conversation Histories & 0.15\% \\
\quad Unauthorized Data Sharing in LLM Chatting & 0.14\% \\
\quad Privacy Risks from File Uploads & 0.12\% \\
\quad Data Storage Risks & 0.11\% \\
\quad Unauthorized Data Aggregation & 0.06\% \\
\midrule
\textbf{Non-compliance with Privacy Regulations} & \textbf{0.55\%} \\
\textit{Perceived failure to adhere to legal privacy frameworks like GDPR.} & \\
\quad General Privacy Compliance Concerns & 0.24\% \\
\quad Ambiguity of Privacy Policy Compliance & 0.23\% \\
\quad Non-compliance with Privacy Regulations & 0.08\% \\
\midrule
\textbf{Privacy Total} & \textbf{3.85\%} \\
\bottomrule
\end{tabular}
\end{table}

\newpage
\begin{table}[htbp]
\centering
\footnotesize
\caption{User-Reported Risks of Fairness}
\label{tab:fairrisks}
\begin{tabular}{p{6.1cm} c}
\toprule
\textbf{Primary Theme / Subtheme Name} & \textbf{Total Share (\%)} \\
\midrule
\textbf{Discriminatory Content Generation} & \textbf{1.36\%} \\
\textit{Produces outputs that stereotype demographic groups or promote ideology.} & \\
\quad Politically Biased Content Moderation & 0.39\% \\
\quad Demographic Stereotyping & 0.37\% \\
\quad Political Bias and Propaganda & 0.24\% \\
\quad Racial Prejudice & 0.20\% \\
\quad Patronizing Tone & 0.16\% \\
\midrule
\textbf{Inequitable Opportunity Creation} & \textbf{1.22\%} \\
\textit{Creates unfair advantages for some and disadvantages for others.} & \\
\quad Inequitable Access & 0.46\% \\
\quad Unfair Market Competition & 0.25\% \\
\quad Enabling Academic Dishonesty & 0.23\% \\
\quad Job Displacement and Wage Disparity & 0.17\% \\
\quad Geographic Restriction & 0.11\% \\
\midrule
\textbf{Bias Propagation through Biased Sources} & \textbf{1.15\%} \\
\textit{Sources cause the LLM to perpetuate unfairness in content generation.} & \\
\quad Subjective Bias & 0.36\% \\
\quad Linguistic Underrepresentation & 0.23\% \\
\quad Bias from Training Data & 0.20\% \\
\quad Reinforcement of Echo Chambers & 0.19\% \\
\quad Erosion of Critical Thinking & 0.17\% \\
\midrule
\textbf{Discriminatory Treatment} & \textbf{0.79\%} \\
\textit{Behaves in ways that treat individuals or groups unfairly.} & \\
\quad Unfairness and Discrimination & 0.20\% \\
\quad Unfair Automated Accusation & 0.19\% \\
\quad Inability to Provide Empathetic Support & 0.16\% \\
\quad Unfair Feature Access & 0.13\% \\
\quad Unequal Recognition & 0.11\% \\
\midrule
\textbf{Fairness Total} & \textbf{4.52\%} \\
\bottomrule
\end{tabular}
\end{table}

\begin{table}[htbp]
\centering
\footnotesize
\caption{User-Reported Risks of Safety (Part 1)}
\label{tab:safetyrisks1}
\begin{tabular}{p{6.1cm} c}
\toprule
\textbf{Primary Theme / Subtheme Name} & \textbf{Total Share (\%)} \\
\midrule
\textbf{Adverse Psychological and Behavioral Influence} & \textbf{2.32\%} \\
\textit{Negatively alters a user's mental state or actions.} & \\
\quad Emotional Reliance Harms & 0.47\% \\
\quad Reinforcement of Negative Behavior & 0.30\% \\
\quad Potential Physical Harm & 0.28\% \\
\quad Social Risk of Emotional Dependency & 0.26\% \\
\quad User Discomfort from Model Behavior & 0.23\% \\
\quad Cognitive Overreliance & 0.18\% \\
\quad Negative Psychological Impact & 0.18\% \\
\quad User Emotional Manipulation & 0.13\% \\
\quad Negative Behavioral Influence & 0.11\% \\
\midrule
\textbf{Inappropriate Content Generation or Moderation} & \textbf{1.60\%} \\
\textit{Creates harmful outputs or fails to moderate them.} & \\
\quad Pornography Content Generation & 0.67\% \\
\quad Toxic Content Generation & 0.57\% \\
\quad Inadequate Content Moderation & 0.14\% \\
\quad Enabling Academic Dishonesty & 0.13\% \\
\quad Intellectual Property Violation & 0.09\% \\
\bottomrule
\end{tabular}
\end{table}

\newpage
\begin{table}[htbp]
\centering
\footnotesize
\caption{User-Reported Risks of Safety (Part 2)}
\label{tab:safetyrisks2}
\begin{tabular}{p{6.1cm} c}
\toprule
\textbf{Primary Theme / Subtheme Name} & \textbf{Total Share (\%)} \\
\midrule
\textbf{Uncontrolled System Behavior towards User Safety} & \textbf{1.46\%} \\
\textit{Acts in a way that compromises a user's online safety.} & \\
\quad Harm from Misusing Personal Information & 0.27\% \\
\quad Safety System Ineffectiveness & 0.25\% \\
\quad Unauthorized Access & 0.25\% \\
\quad Threatening Behavior towards Users & 0.20\% \\
\quad Unintended Autonomous Behavior & 0.20\% \\
\quad Uncalibrated Refusal Behavior & 0.19\% \\
\quad Harm from Usage Issues & 0.10\% \\
\midrule
\textbf{Dangerous Advice Provision} & \textbf{0.80\%} \\
\textit{Dispenses advice posing a direct risk to user health/safety.} & \\
\quad Dangerous Instructions & 0.33\% \\
\quad Inappropriate Therapeutic Advice & 0.16\% \\
\quad Harmful Medical Advice & 0.16\% \\
\quad Generation of Dangerous Instructions & 0.15\% \\
\midrule
\textbf{Negative Environmental Impact} & \textbf{0.20\%} \\
\textit{Damages the environment via high resource consumption.} & \\
\quad Environmental Harm from Resource Consumption & 0.20\% \\
\midrule
\textbf{Safety Total} & \textbf{6.39\%} \\
\bottomrule
\end{tabular}
\end{table}

\begin{table}[htbp]
\centering
\footnotesize
\caption{User-Reported Risks of Security \& Resilience}
\label{tab:securerisks}
\begin{tabular}{p{6.1cm} c}
\toprule
\textbf{Primary Theme / Subtheme Name} & \textbf{Total Share (\%)} \\
\midrule
\textbf{Exploitation Through Security Vulnerabilities} & \textbf{3.51\%} \\
\textit{Be compromised through malicious use or inherent weaknesses.} & \\
\quad Jailbreaking Vulnerabilities & 0.83\% \\
\quad Prompt Injection Attacks & 0.55\% \\
\quad Data Leakage & 0.47\% \\
\quad Manipulative Interaction & 0.40\% \\
\quad Cybersecurity Risks & 0.35\% \\
\quad Unauthorized Intrusion & 0.33\% \\
\quad Detection Evasion & 0.22\% \\
\quad API Authentication Vulnerabilities & 0.16\% \\
\quad DDoS Attacks & 0.10\% \\
\quad Account Suspension & 0.08\% \\
\midrule
\textbf{Operational Instability and Disruption} & \textbf{3.18\%} \\
\textit{Failure to maintain consistent and stable function during use.} & \\
\quad Downtime \& Maintenance & 1.51\% \\
\quad Service Disruption & 0.44\% \\
\quad Performance Bottlenecks & 0.40\% \\
\quad Service Latency & 0.25\% \\
\quad Feature Malfunctions & 0.15\% \\
\quad System Downtime & 0.15\% \\
\quad Backup Failures & 0.15\% \\
\quad Session Interruptions & 0.14\% \\
\midrule
\textbf{Infrastructure and Resource Limitations} & \textbf{2.59\%} \\
\textit{Underlying architectural constraints that threaten performance.} & \\
\quad Usage Limitation & 0.57\% \\
\quad Rate Limit Risks & 0.45\% \\
\quad Workload Overload & 0.31\% \\
\quad Unauthorized Replication of Models & 0.27\% \\
\quad Memory Overload & 0.22\% \\
\quad Token Limitation Risks & 0.21\% \\
\quad Access Restriction Risks & 0.21\% \\
\quad Moderation Restrictions & 0.17\% \\
\quad Regional Restrictions & 0.11\% \\
\quad Costly API Usage & 0.07\% \\
\midrule
\textbf{Security and Resilience Total} & \textbf{9.27\%} \\
\bottomrule
\end{tabular}
\end{table}

\newpage
\begin{table}[htbp]
\centering
\footnotesize
\caption{User-Reported Risks of Accountability \& Transparency}
\label{tab:accountabilityrisks}
\begin{tabular}{p{6.1cm} c}
\toprule
\textbf{Primary Theme / Subtheme Name} & \textbf{Total Share (\%)} \\
\midrule
\textbf{Unclear Operational Policies and Practices} & \textbf{3.81\%} \\
\textit{Lack of transparency in the rules governing operational behavior.} & \\
\quad Operational Opacity & 1.17\% \\
\quad Opaque Access Restrictions & 0.52\% \\
\quad Unclear Usage Limitations & 0.50\% \\
\quad Deceptive or Manipulative Behavior & 0.50\% \\
\quad Opaque Content Moderation & 0.42\% \\
\quad Opaque Engagement-Driven Behavior & 0.41\% \\
\quad Facilitating Dishonesty & 0.28\% \\
\midrule
\textbf{Obscured Product Specification} & \textbf{3.46\%} \\
\textit{Lack of clear information about what the LLM is and can do.} & \\
\quad Obscured Functionality Through Poor Usability & 0.66\% \\
\quad Model Capability Ambiguity & 0.62\% \\
\quad Opaque Model Versioning & 0.54\% \\
\quad Lack of User Control & 0.42\% \\
\quad Unclear Feature Availability & 0.34\% \\
\quad Deceptive Anthropomorphism & 0.32\% \\
\quad Lack of General Transparency Features & 0.31\% \\
\quad Misleading Product Labeling & 0.25\% \\
\midrule
\textbf{Opaque Commercial Practices} & \textbf{2.76\%} \\
\textit{Lack of transparency in the business models surrounding the LLM.} & \\
\quad Unclear Pricing & 0.62\% \\
\quad Inadequate Disclosure of Information to the Users & 0.41\% \\
\quad Opaque Subscription and Billing & 0.41\% \\
\quad Deceptive Marketing & 0.40\% \\
\quad Opaque Market Influence & 0.37\% \\
\quad Opaque Revenue Models & 0.28\% \\
\quad Opaque Monetization Incentives & 0.27\% \\
\midrule
\textbf{Absence of High-level Product Governance} & \textbf{2.42\%} \\
\textit{Lack of clear high-level structures for oversight.} & \\
\quad Lack of Accountability for AI Actions & 0.67\% \\
\quad Lack of Clear Accountability Structures & 0.47\% \\
\quad Insufficient Guiding Documentation & 0.46\% \\
\quad Unclear Intellectual Property Liability & 0.38\% \\
\quad Lack of Ethical Oversight & 0.28\% \\
\quad Lack of Regulatory Compliance Transparency & 0.15\% \\
\midrule
\textbf{Opaque Information Processing} & \textbf{2.10\%} \\
\textit{The ``black box'' nature of how the LLM was trained or processes info.} & \\
\quad Lack of Data Source Transparency & 0.66\% \\
\quad Unclear Instruction Processing & 0.49\% \\
\quad Opaque Context and Memory Management & 0.41\% \\
\quad Unpredictable Autonomous Behavior & 0.25\% \\
\quad Opaque Training Data and Processes & 0.20\% \\
\quad Obscured Prompt and Output Processing & 0.10\% \\
\midrule
\textbf{Failure of Post-Interaction Redress} & \textbf{1.04\%} \\
\textit{Breakdown in accountability after a negative user outcome.} & \\
\quad Inadequate Feedback Channels & 0.37\% \\
\quad Refusal to Acknowledge Errors & 0.37\% \\
\quad Erosion of User Trust & 0.30\% \\
\midrule
\textbf{Obscured Content Origin} & \textbf{0.76\%} \\
\textit{Inability to determine the source of content from the LLM.} & \\
\quad Blurring of Human-AI Interaction & 0.36\% \\
\quad Opaque Origin of Generated Content & 0.21\% \\
\quad Lack of Authorship Transparency & 0.19\% \\
\midrule
\textbf{Accountability and Transparency Total} & \textbf{16.35\%} \\
\bottomrule
\end{tabular}
\end{table}

\newpage
\begin{table}[htbp]
\centering
\footnotesize
\caption{User-Reported Risks of Validity \& Reliability}
\label{tab:validrisks}
\begin{tabular}{p{6.1cm} c}
\toprule
\textbf{Primary Theme / Subtheme Name} & \textbf{Total Share (\%)} \\
\midrule
\textbf{Consistency and Reliability Failures} & \textbf{12.12\%} \\
\textit{Inconsistent, unreliable, or unpredictable outputs.} & \\
\quad Inefficient Usability Performance & 2.56\% \\
\quad Output Inconsistency & 1.34\% \\
\quad Inconsistent Responses & 1.30\% \\
\quad Model-Specific Inconsistencies & 1.29\% \\
\quad Output Unpredictability and Unreliability & 1.17\% \\
\quad Inaccurate and Inconsistencies & 1.12\% \\
\quad Inconsistent Outputs & 1.08\% \\
\quad Output Formatting Inconsistency & 0.89\% \\
\quad Output Uncertainty & 0.61\% \\
\quad Input Quality Sensitivity & 0.58\% \\
\quad Performance Failures & 0.18\% \\
\midrule
\textbf{Factual Invalidity of Content Generation} & \textbf{10.96\%} \\
\textit{Outputs deviate from facts, mislead users, or fabricate details.} & \\
\quad Hallucinations & 3.95\% \\
\quad Misinformation and Misrepresentation & 2.94\% \\
\quad Factual Inaccuracy & 1.56\% \\
\quad Capability Overestimation & 1.08\% \\
\quad Overconfident Outputs & 0.80\% \\
\quad Overreliance and Overfitting & 0.64\% \\
\midrule
\textbf{Modality Gaps in Content Generation} & \textbf{7.04\%} \\
\textit{Specialized content generation capabilities are inadequate.} & \\
\quad Inefficient Code Generation & 1.94\% \\
\quad Subpar Image Quality & 1.28\% \\
\quad Lack of Writing Originality & 1.19\% \\
\quad Repetitive Content Generation & 0.85\% \\
\quad Multilingual Inaccuracies & 0.82\% \\
\quad Ineffective Web Search & 0.63\% \\
\quad Audio Processing Issues & 0.33\% \\
\midrule
\textbf{Context Understanding or Memory Limits} & \textbf{6.65\%} \\
\textit{Context retention fails and conversational history is not maintained reliably.} & \\
\quad Context Window Limitations & 1.89\% \\
\quad Poor Reasoning & 1.34\% \\
\quad Memory and Recall Failures & 1.26\% \\
\quad Context-Dependent Inaccuracies & 1.11\% \\
\quad Conversational Context Failures & 1.05\% \\
\midrule
\textbf{Inherent Training or Architecture Limitations} & \textbf{6.41\%} \\
\textit{Inherent design constraints limit what the model can process.} & \\
\quad Hardware Performance Bottlenecks & 2.55\% \\
\quad Inefficient Cost Performance & 1.38\% \\
\quad Training Data Issues & 0.97\% \\
\quad Tokenization Limitations & 0.81\% \\
\quad Input/Output Length Limitation & 0.70\% \\
\midrule
\textbf{Benchmark Underperformance or Performance Degradation} & \textbf{6.06\%} \\
\textit{Speed, accuracy, or efficiency lags baselines, or degrades.} & \\
\quad Performance Degradation Over Time & 2.19\% \\
\quad Performance Degradation & 1.83\% \\
\quad Underperformance in Benchmarks & 1.09\% \\
\quad Poor Code Bench Performance & 0.51\% \\
\quad Degradation from Quantization & 0.44\% \\
\midrule
\textbf{Software Functionality Failures} & \textbf{5.55\%} \\
\textit{Core functions break, features malfunction, or error loops recur.} & \\
\quad Functional Limitations & 1.84\% \\
\quad General Malfunctions and Inconsistencies & 1.30\% \\
\quad Feature Malfunctions & 0.93\% \\
\quad Repetitive Loops and Errors & 0.86\% \\
\quad Usage Limit and Service Disruption & 0.62\% \\
\midrule
\textbf{Task Comprehension Failures} & \textbf{3.60\%} \\
\textit{Misreads tasks, misinterprets guides, or fails to execute requests.} & \\
\quad Prompt Misinterpretation & 1.27\% \\
\quad Instruction Non-Adherence & 0.84\% \\
\quad Inefficient Large-Scale Input Processing & 0.76\% \\
\quad User Intent Misinterpretation & 0.73\% \\
\midrule
\textbf{Validity and Reliability Total} & \textbf{58.39\%} \\
\bottomrule
\end{tabular}
\end{table}

\end{document}